\documentclass[aps,prb,twocolumn,groupedaddress]{revtex4}

\usepackage{graphicx}
\usepackage{dcolumn}
\usepackage{bm}
\usepackage{amssymb}
\usepackage{color}

\begin{document}


\title{Structure formation in binary mixtures of lipids and detergents:
Self-assembly and vesicle division
}

\author{Hiroshi Noguchi}
\email[]{noguchi@issp.u-tokyo.ac.jp}
\affiliation{
Institute for Solid State Physics, University of Tokyo,
 Kashiwa, Chiba 277-8581, Japan}

\date{\today}

\begin{abstract}
Self-assembly dynamics in binary surfactant mixtures and
structure changes of lipid vesicles induced by detergent solution
are studied using coarse-grained molecular simulations.
Disk-shaped micelles, the bicelles, are stabilized by  detergents surrounding the rim of a bilayer disk of lipids.
The self-assembled bicelles are considerably smaller than bicelles formed from vesicle rupture,
and their size is determined by the concentrations of lipids and detergents and the interactions between the two species.
The detergent-adsorption induces spontaneous curvature of the vesicle bilayer
and results in vesicle division into two vesicles or vesicle rupture into worm-like micelles.
The division occurs mainly via the inverse pathway of the modified stalk model.
For large spontaneous curvature of the monolayers of the detergents, a pore is often opened, thereby leading to
 vesicle division or worm-like micelle formation.
\end{abstract}


\maketitle

\section{Introduction}

Amphiphilic molecules such as lipids and detergents
self-assemble into various structures in aqueous solutions \cite{isra11,safr94,jain03,sedd04,caff09,ujwa11,walt91,yama09,jain04}.
A single type of simple surfactant can
form spherical and cylindrical (worm-like) micelles, bilayer membranes, inverted hexagonal structures, and inverted micelles
depending on the relative size of their hydrophilic parts \cite{isra11,safr94}.
Aggregates of different types of surfactants
can exhibit more complicated self-assembled structures such as undulated cylinders, octopus-like micelles, and vesicles connected with
worm-like micelles \cite{jain04,chri09}.
In mixtures of lipids and cone-shaped surfactants with a relatively large hydrophilic head,
the edges of bilayer membranes are stabilized by surfactants, and
a bilayer disk called bicelle is formed \cite{sedd04,caff09,ujwa11,walt91,yama09}.
The bicelle is an excellent medium to investigate protein functions,
since membrane proteins can maintain their functionality in the bicelles \cite{sand98,ujwa11}. 
Bicelles can align in magnetic fields and are used for NMR studies \cite{ange07}.
The structures of several proteins have been determined by the bicelle method \cite{ujwa11,faha02}.

The dynamics of micelle growth and subsequent vesicle formation
 have been studied experimentally by time-resolved scatting techniques (light, neutron, 
and X-ray) \cite{leng02,weis05,brys05,made11,gumm11}.
Mixtures of lipids (lecithin) and  natural detergents (bile salts) \cite{leng02,made11} or
cationic and anionic surfactants \cite{weis05,gumm11}
are often used in experiments, since the size and shape of the surfactant aggregates can be changed by the concentration variation.
Bile salts are also typical detergents that are used to produce bicelles.
The self-assembly of lipids into vesicles in single-component lipid systems has been
studied using simulations of a meshless membrane model with and without hydrodynamics \cite{nogu06a}.
However, in contrast, the self-assembly dynamics of lipids and detergents to form bicelles has not been completely understood.
The first aim of this study is to investigate the self-assembly dynamics of binary mixtures of lipids and detergents.

When a detergent solution is added to a liposome suspension,
the liposomes can dissolve.
Experimental studies have revealed that
liposomes take
various types of solubilization pathways  depending on 
the types of lipids and detergents \cite{nomu01,stan05,sudb11,tomi11,elsa11,hama12};
these including rhythmic shrinkage, bursting, budding, fission, peeling, and inside-out inversion.
In particular,  vesicle fission (or division) is considered an important dynamics 
since vesicle growth and division are two fundamental processes of the self-reproduction.
Recently, the self-reproduction of vesicles 
has received growing attention in terms of both
experiments \cite{szos01,hanc04,zepi08,kuri11,tera12} and theory \cite{svet09,bozi07},
since it is one of the key functions of protocells. 
To our knowledge, surfactant-adsorption-induced vesicle division has been numerically investigated
 only in a few simulations carried out by Markvoort and coworkers \cite{mark10}. 
The second aim of this study is to systematically investigate the dynamics 
and mechanism of detergent-adsorption-induced vesicle division.

Membrane fusion and fission are  key events in various intra- and intercellular
processes in living cells,
and they have been intensively investigated \cite{jahn02,cher08,niko11,mark11,mull11}.
The fission dynamics of a bilayer membrane has been simulated for single-component \cite{nogu02a,nogu02b,nogu03}
 and two-component membranes  \cite{yama03,smit07,mark10}.
Membrane fission occurs via pathways opposite to those of membrane fusion
via a stalk intermediate.
In one of the fission pathways, the formation of a trans-monolayer contact \cite{sieg93} induces the stalk formation\cite{nogu02b,mark10}, and
in the other, stalk formation is induced by opening of a small pore \cite{nogu02a,yama03}.
The control parameters that determine these fission and fusion pathways have not been well understood so far.

In order to simulate bilayer membranes on longer and larger scales,
various types of coarse-grained molecular models have been proposed (see review articles \cite{muel06,vent06,nogu09,marr09,shin12}).
In the present study,  we employ  one of the solvent-free molecular models \cite{nogu11}.
In this model, membrane properties such as the bending rigidity and spontaneous curvature of the monolayer
 can be varied over wide ranges.
In our previous study \cite{nogu12a}, 
we found that the rupture of two-component vesicles can lead to the formation of variously shaped micelles and vesicles including
bicelles and octopus-like micelles.
We apply the same model to the molecular self-assembly into bicelles and detergent adsorption to vesicles in this paper.

In Sec. \ref{sec:method}, the surfactant model and simulation method are briefly described.
In Sec. \ref{sec:ass}, self-assembly from isolated molecules into bicelles and vesicles
are investigated.
In  Sec. \ref{sec:abs}, detergent-adsorption-induced topological changes, vesicle division and rupture into worm-like micelles 
are described.
The summary is provided in Sec. \ref{sec:sum}.

\section{Model and Method}
\label{sec:method}

We employ a solvent-free molecular model  \cite{nogu11,nogu12a} to simulate two-component surfactant mixtures.
One type (type A) of molecules  is considered as the detergent molecule, being conically shaped with a relatively
large head (the spontaneous curvature of the monolayer  $C_0 \ge 0$).
The other type (type B) is considered as a lipid with a cylindrical shape ($C_0 =0$).
The surfactant model and simulation method are only briefly described here, since
the details of the model can be found in Refs. \onlinecite{nogu11,nogu12a}. 

Each ($i$-th) molecule of the types A and B has a spherical particle with an orientation vector ${\bf u}_i$, 
which represents the direction from the hydrophobic to the hydrophilic part.
There are two points of interaction in the molecule:
the center of a sphere ${\bf r}^{\rm s}_i$ and a hydrophilic point ${\bf r}^{\rm e}_i={\bf r}^{\rm s}_i+{\bf u}_i \sigma$.
Type A molecules are considered for $0 \le i< N_{\rm A}$ and type B for  $N_{\rm A} \le i< N=N_{\rm A}+N_{\rm B}$.
The molecules interact with each other via the potential given by
\begin{eqnarray}
\frac{U}{k_{\rm B}T} &=\ \ & \hspace{1cm} \sum_{i<j} \exp[-20(r_{ij}^{\rm s}/\sigma-1)]  \label{eq:U_all}
               + \sum_{i} \varepsilon_i\  U_{\rm {att}}(\rho_i) \nonumber \\ \nonumber
&\ \ +& \ \ \frac{k_{\rm{tilt}}}{2} \sum_{i<j} \bigg[ 
( {\bf u}_{i}\cdot \hat{\bf r}^{\rm s}_{ij})^2
 + ({\bf u}_{j}\cdot \hat{\bf r}^{\rm s}_{ij})^2  \bigg] w_{\rm {cv}}(r^{\rm e}_{ij}) \\ \nonumber
&\ \ +&  \frac{k_{\rm {bend}}}{2} \sum_{i<j}  \bigg({\bf u}_{i} - {\bf u}_{j} - C_{\rm {bd}}^{ij} \hat{\bf r}^{\rm s}_{ij} \bigg)^2 w_{\rm {cv}}(r^{\rm e}_{ij}) \\ 
&\ \ +&   \varepsilon_{\rm {AB}} \sum_{i < N_{\rm A}, j \ge N_{\rm A}} U_{\rm {AB}}(r_{ij}^{\rm s}),
\end{eqnarray} 
where ${\bf r}_{ij} = {\bf r}_{i}-{\bf r}_j$, $r_{ij} = |{\bf r}_{ij}|$,
 $\hat{\bf r}_{ij}={\bf r}_{ij}/r_{ij}$, and $k_{\rm B}T$ denotes the thermal energy.

An excluded volume of the molecules with a diameter $\sigma$
and an attractive interaction between the molecules are given by the first and second terms in Eq. (\ref{eq:U_all}), respectively.
The two types of molecules can have different attractive strengths:
$\varepsilon_i = \varepsilon_{\rm att}^{\rm A}$  for $0 \le i< N_{\rm A}$ 
and $\varepsilon_i = \varepsilon_{\rm att}^{\rm B}$  for $N_{\rm A} \le i< N$.
The potential $U_{\rm {att}}(\rho_i)$ is given by
\begin{equation} \label{eq:U_att}
U_{\rm {att}}(\rho_i) = 0.25\ln[1+\exp\{-4(\rho_i-\rho^*)\}]- C,
\end{equation}
with $C= 0.25\ln\{1+\exp(4\rho^*)\}$.
The local particle density $\rho_i$ is approximately the number of
particles ${\bf r}^{\rm s}_i$  in the sphere with radius $r_{\rm {att}}$,
\begin{equation}
\rho_i= \sum_{j \ne i} f_{\rm {cut}}(r^{\rm s}_{ij}), 
\label{eq:wrho}
\end{equation} 
where $f_{\rm {cut}}(r)$ is a $C^{\infty}$ cutoff function~\cite{nogu06},
\begin{equation} \label{eq:cutoff}
f_{\rm {cut}}(r)=\left\{ 
\begin{array}{ll}
\exp\{A_0(1+\frac{1}{(r/r_{\rm {cut}})^n -1})\}
& (r < r_{\rm {cut}}) \\
0  & (r \ge r_{\rm {cut}}) 
\end{array}
\right.
\end{equation}
with $n=6$, $A_0=\ln(2) \{(r_{\rm {cut}}/r_{\rm {att}})^n-1\}$,
$r_{\rm {att}}= 1.9\sigma$, 
and the cutoff radius $r_{\rm {cut}}=2.4\sigma$.

The third and fourth terms in Eq.~(\ref{eq:U_all}) are
discretized versions of the 
tilt and bending potentials of the tilt model \cite{hamm98,hamm00}, respectively.
A smoothly truncated Gaussian function~\cite{nogu06} 
is employed as the weight function 
\begin{equation} \label{eq:wcv}
w_{\rm {cv}}(r)=\left\{ 
\begin{array}{ll}
\exp (\frac{(r/r_{\rm {ga}})^2}{(r/r_{\rm {cc}})^n -1})
& (r < r_{\rm {cc}}) \\
0  & (r \ge r_{\rm {cc}}) 
\end{array}
\right.
\end{equation}
with  $n=4$, $r_{\rm {ga}}=1.5\sigma$, and $r_{\rm {cc}}=3\sigma$.
The spontaneous curvatures of the monolayer membranes
are varied by parameters $C_{\rm {bd}}^{\rm A}$ and $C_{\rm {bd}}^{\rm B}$:
$C_{\rm {bd}}^{ij} =  (C_{\rm {bd}}^{i}+ C_{\rm {bd}}^{j})/2$,
 $C_{\rm {bd}}^{i} = C_{\rm {bd}}^{\rm A}$  for $0 \le i< N_{\rm A}$, 
and $C_{\rm {bd}}^{i} = C_{\rm {bd}}^{\rm B}$  for $N_{\rm A} \le i< N$.

The last term in Eq.~(\ref{eq:U_all}) represents the repulsion between
the different types of molecules
as a monotonic decreasing function:
$U_{\rm {AB}}(r) =  A_1 f_{\rm {cut}}(r)$ with $n=1$, $A_0= 1$, $r_{\rm {cut}}=2.4\sigma$,
and $A_1=\exp[\sigma/(r_{\rm {cut}}-\sigma)]$.

The  $NVT$ ensemble (constant
number of molecules $N$, volume $V$, and temperature $T$) is used
with periodic boundary conditions in a cubic box with side length $L$.
Underdamped Langevin equations [Brownian dynamics (BD)]
are used for time evolution of the molecules.
The motion of the center of the mass 
${\bf r}^{\rm G}_{i}=({\bf r}^{\rm s}_{i}+{\bf r}^{\rm e}_{i})/2 $ and 
the orientation ${\bf u}_{i}$ are given by
\begin{eqnarray} \label{eq:lan1}
  \frac{d {\bf r}^{\rm G}_{i}}{dt} &=& {\bf v}^{\rm G}_{i}, \ \  \frac{d {\bf u}_{i}}{dt} = {\boldsymbol \omega}_{i}, \\
m \frac{d {\bf v}^{\rm G}_{i}}{dt} &=&
 - \zeta_{\rm G} {\bf v}^{\rm G}_{i} + {\bf g}^{\rm G}_{i}(t)
 + {{\bf f}_i}^{\rm G}, \label{eq:lan2}  \\ \label{eq:lan3}
I \frac{d {\boldsymbol \omega}_{i}}{dt} &=&
 - \zeta_{\rm r} {\boldsymbol \omega}_i + ({\bf g}^{\rm r}_{i}(t)
 + {{\bf f}_i}^{\rm r})^{\perp} + \lambda_{\rm L} {\bf u}_{i},
\end{eqnarray}
where $m$ and $I$ denote the mass and the moment of inertia of the molecule, respectively.
The forces are given by ${{\bf f}_i}^{\rm G}= - \partial U/\partial {\bf r}^{\rm G}_{i}$
and ${{\bf f}_i}^{\rm r}= - \partial U/\partial {\bf u}_{i}$ with 
the perpendicular component ${\bf a}^{\perp} ={\bf a}- ({\bf a}\cdot{\bf u}_{i}) {\bf u}_{i}$,
and a Lagrange multiplier $\lambda_{\rm L}$ is used to ensure that ${\bf u}_{i}^2=1$.
The friction coefficients $\zeta_{\rm G}$ and $\zeta_{\rm r}$ and 
the Gaussian white noises ${\bf g}^{\rm G}_{i}(t)$ and ${\bf g}^{\rm r}_{i}(t)$
obey the fluctuation-dissipation theorem.

\begin{figure}
\includegraphics{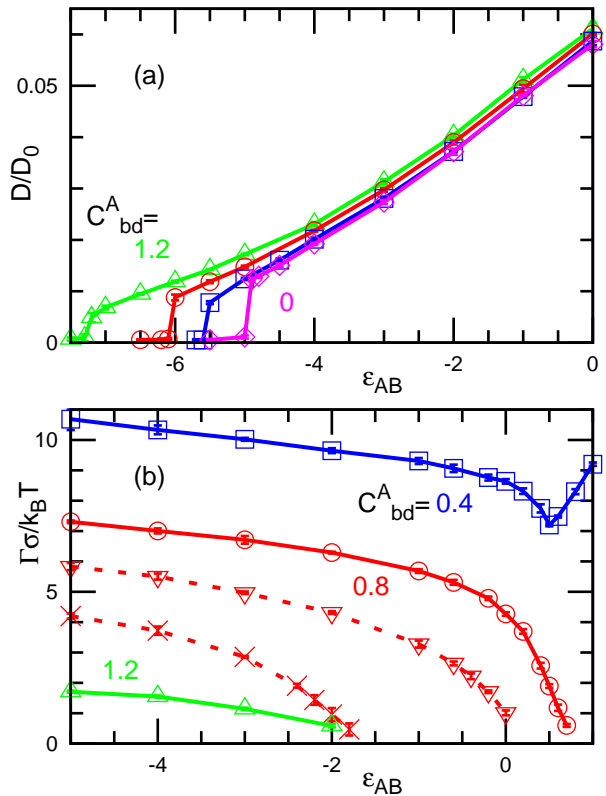}
\caption{\label{fig:dif}
Dependence of
(a) the lateral diffusion coefficient $D$ and (b) the effective line tension $\Gamma$ of the membrane edges on $\varepsilon_{\rm {AB}}$ 
at $ N_{\rm A}=N_{\rm B}$.
(a) The symbols with lines represent
data for $C_{\rm {bd}}^{\rm {A}}=0$ ($\diamond$), $0.4$ ($\Box$), $0.8$ ($\circ$), and $1.2$ ($\triangle$)
at $\varepsilon_{\rm att}^{\rm A}=2$.
The diffusion coefficient $D$ is normalized using the diffusion coefficient $D_0=\sigma^2/\tau_0$ of an isolated molecule.
(b) The symbols with solid lines represent data at $\varepsilon_{\rm att}^{\rm A}=2$.
The symbols with dashed lines  represent data at $\varepsilon_{\rm att}^{\rm A}=1.5$ ($\triangledown$) and $1$ ($\times$).
}
\end{figure}

In this paper, 
we use $k_{\rm {bend}}=k_{\rm {tilt}}=8$, $\rho^*=14$,  $\varepsilon_{\rm {att}}^{\rm B}=2$, and  $C_{\rm bd}^{\rm B}=0$.
The type B (lipid) molecules form a fluid bilayer membrane with $\kappa/k_{\rm B}T=44$,
and they have very low values of CMC. 
No molecules are detached from a single-component membrane of type B in typical simulation time scales.
For the type A (detergent) molecules, the attraction range is set to $0.6 \leq \varepsilon_{\rm {att}}^{\rm A} \leq 1.5$
so that these molecules have a finite CMC range $\sim 10^{-4}/\sigma^{3}$ to $10^{-2}/\sigma^{3}$.
The spontaneous curvature $C_0$ can be varied by the parameter $C_{\rm {bd}}$ as
$C_0  \simeq \{k_{\rm {bend}}/(k_{\rm {bend}}+k_{\rm {tilt}})\}C_{\rm {bd}}/\sigma$. \cite{shiba11,nogu12}
At $k_{\rm {bend}}=k_{\rm {tilt}}$,  $C_0  \simeq C_{\rm {bd}}/2\sigma$.
For type A molecules, the curvature parameter range is set to  $C_{\rm {bd}}^{\rm A}=0$ to $1.2$
($C_0^{\rm A}=0$ to $0.6\sigma$).
A single-component vesicle is ruptured into worm-like micelles
when $C_0 \gtrsim 0.8$ at $\varepsilon_{\rm {att}}=2$. \cite{nogu11}

In order to investigate molecular self-assembly (Sec. \ref{sec:ass}), we set
randomly distributed gas states at $\varepsilon_{\rm {att}}^{\rm A}=\varepsilon_{\rm {att}}^{\rm B}=0.5$
and $C_{\rm {bd}}^{\rm A}=0$ as initial states
with  $N=N_{\rm A}+N_{\rm B}=4,000$ and $L=40\sigma$ (the total concentration $\rho=N/V=0.0625/\sigma^3$).
Subsequently, the simulations are begun
with $\varepsilon_{\rm {att}}^{\rm A}=1.5$ and $C_{\rm {bd}}^{\rm A}=0.8$ ($C_0^{\rm A}=0.4\sigma$).

The repulsion between type A and type B molecules is varied
from $\varepsilon_{\rm {AB}}=0$ to $3$.
In tensionless membranes,
the type A and B molecules are phase-separated at $\varepsilon_{\rm {AB}} \gtrsim 1$,
and the line tension $\Gamma_{\rm {AB}}$ between the type A and type B domains increases with increasing $\varepsilon_{\rm {AB}}$:
$\Gamma_{\rm {AB}}=1.7$ and $3.5$ at $\varepsilon_{\rm {AB}}=1.5$ and $2$, respectively.
At $\varepsilon_{\rm {AB}} \gtrsim 2$, flat membranes become unstable;
the membrane is ruptured from the domain boundary,
and subsequently, type A molecules form worm-like micelles.

In order to investigate surfactant absorption into vesicles (Sec. \ref{sec:abs}),
 the type A molecules are distributed outside of 
a vesicle composed of type B molecules for  $N_{\rm A}=8,000$, $N_{\rm B}=2,000$, and $L=80\sigma$.
The concentrations of the type A and B molecules are given by $\rho_{\rm A}=0.0156/\sigma^3$ and $\rho_{\rm B}=0.0039/\sigma^3$, respectively.
Before the execution of the product runs, the molecules are equilibriated
 at  $\varepsilon_{\rm {AB}} = 10$ so that no type A molecules are absorbed into the vesicle.
At the beginning of the product runs ($t=0$),  $\varepsilon_{\rm {AB}}$ is changed to a  negative value.
The negative values of  $\varepsilon_{\rm {AB}}$ are used to generate attraction between type A and type B molecules.
Positively-charged surfactants and negatively-charged surfactants can have similar attraction, and
this attraction enhances absorption of the type A molecules into the vesicle.

The results are displayed with a length unit of $\sigma$, an energy unit of $k_{\rm B}T$, and
 a time unit of $\tau_0=\zeta_{\rm G}\sigma^2/k_{\rm B}T$.
The error bars of the data are estimated 
from the standard errors of ten independent runs.
The time unit is estimated as $\tau_{\rm 0} \sim 0.1 \mu$s from $\sigma \simeq 2$nm and
the lateral diffusion coefficient $\sim 10^{-8}$cm$^2$/s for phospholipids \cite{wu77}.

Figure \ref{fig:dif} shows the  $\varepsilon_{\rm {AB}}$ dependence of the lateral diffusion coefficient $D$ and the effective line tension $\Gamma$ of the membrane edges.
The quantities $D$ and $\Gamma$ are estimated from the mean square displacement and pressure tensor for $N_{\rm A}=N_{\rm B}$, respectively \cite{nogu12a}.
As  $\varepsilon_{\rm {AB}}$ decreases, $D$ decreases and subsequently exhibits a jump into zero at $\varepsilon_{\rm {AB}} \sim -6$.
At $\varepsilon_{\rm {AB}} \lesssim -6$,
the membrane is in gel phase.
Here, we mainly use the parameter region of fluid membranes.
The line tension $\Gamma$ of the edges of two-component membranes
depends on $\varepsilon_{\rm {AB}}$, since the positive spontaneous curvature of type A molecules can reduce the 
free energy of the edges.
As $\varepsilon_{\rm {AB}}$ increases,
 $\Gamma$ decreases and reaches
 a  minimum value at the phase separation point.
For completely phase-separated membranes, $\Gamma$  increases as the sum of the two line tensions: 
$\Gamma=\Gamma_{\rm {A}}+\Gamma_{\rm {AB}}$ \cite{nogu12a}  [see the blue line with squares in Fig. \ref{fig:dif}(b)].
With decreasing $\varepsilon_{\rm att}^{\rm A}$ at $C_{\rm {bd}}^{\rm A}=0.8$, 
the line tension $\Gamma$ decreases, and subsequently, the bilayer edges become unstable, thereby leading to the formation of worm-like micelles.

\section{Self-assembly into bicelles}
\label{sec:ass}

\begin{figure}
\includegraphics{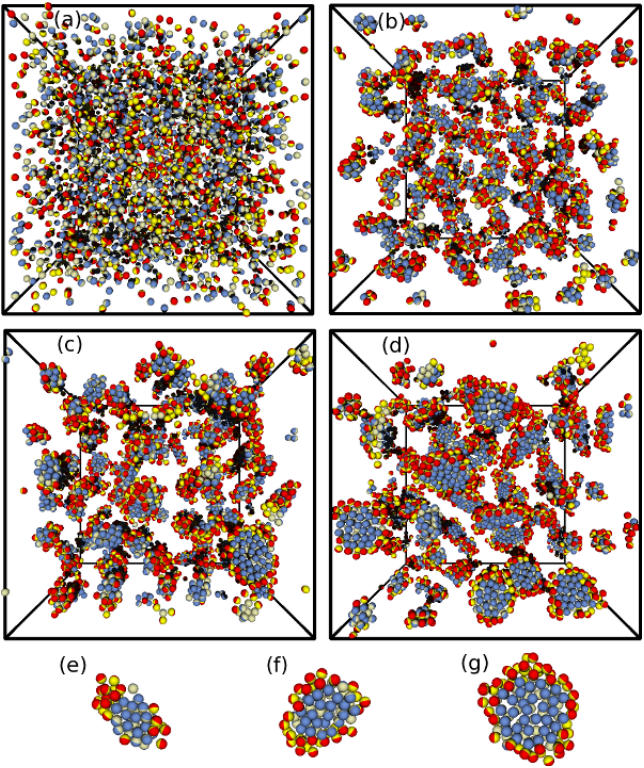}
\caption{\label{fig:ass_snap_r1}
Sequential snapshots of self-assembly into bicelles
at $N_{\rm A}=N_{\rm B}=2,000$ and $\varepsilon_{\rm {AB}}=1$.
(a-d) All molecules in the simulation are shown at 
(a) $t=0$, (b) $1,000\tau_0$, (c) $100,000\tau_0$, and (d) $1,000,000\tau_0$.
(e-g) The largest cluster is shown at (e)  $t=1,000\tau_0$, (f) $10,000\tau_0$, and (g) $1,000,000\tau_0$. 
The red (yellow) and light blue (light yellow) hemispheres represent the hydrophilic 
(hydrophobic) parts of type A  and type B molecules,
respectively.
}
\end{figure}

\begin{figure}
\includegraphics{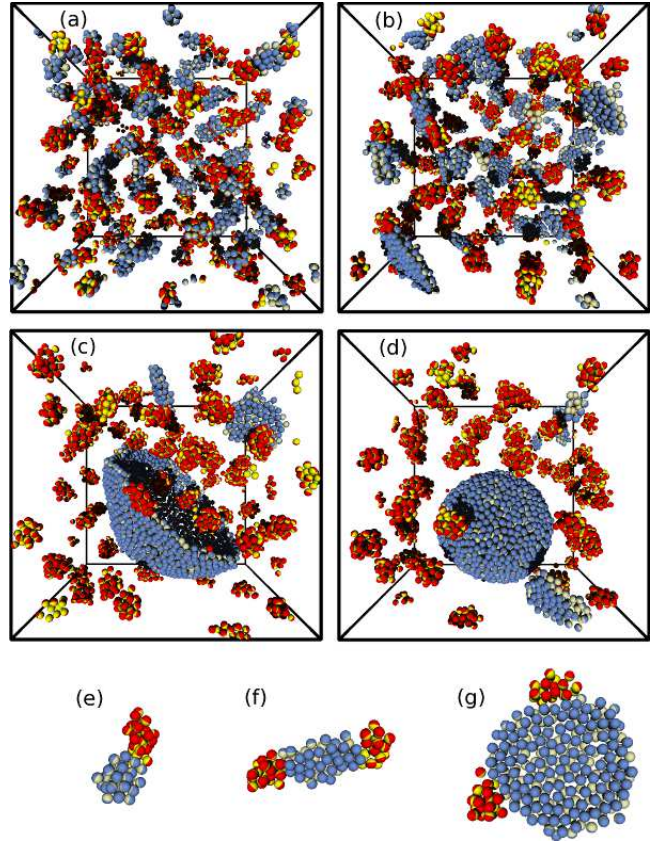}
\caption{\label{fig:ass_snap_r3}
Sequential snapshots of self-assembly into a vesicle and micelles
at $N_{\rm A}=N_{\rm B}=2,000$ and $\varepsilon_{\rm {AB}}=3$.
(a-d) All molecules in the simulation are shown at (a) $t=5,000\tau_0$, (b) $100,000\tau_0$, (c) $803,000\tau_0$, and (d) $1,000,000\tau_0$.
(e-g) The largest cluster is shown at (e)  $t=5,000\tau_0$, (f) $30,000\tau_0$, and (g) $100,000\tau_0$. 
}
\end{figure}

\begin{figure}
\includegraphics{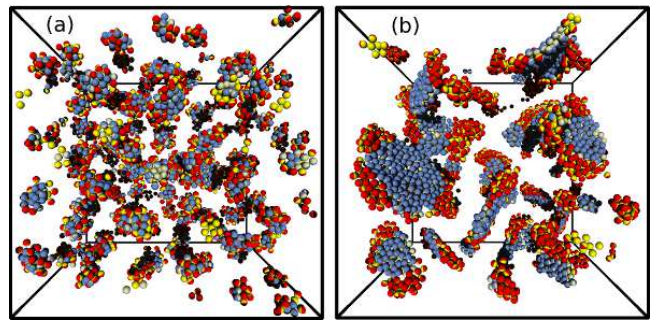}
\caption{\label{fig:ass_snap_r02}
Snapshots of self-assembled micelles 
for (a)  $\varepsilon_{\rm {AB}}=0$ and (b)  $\varepsilon_{\rm {AB}}=2$
at $N_{\rm A}=N_{\rm B}=2,000$ and $t=1,000,000\tau_0$.
}
\end{figure}

\begin{figure}
\includegraphics{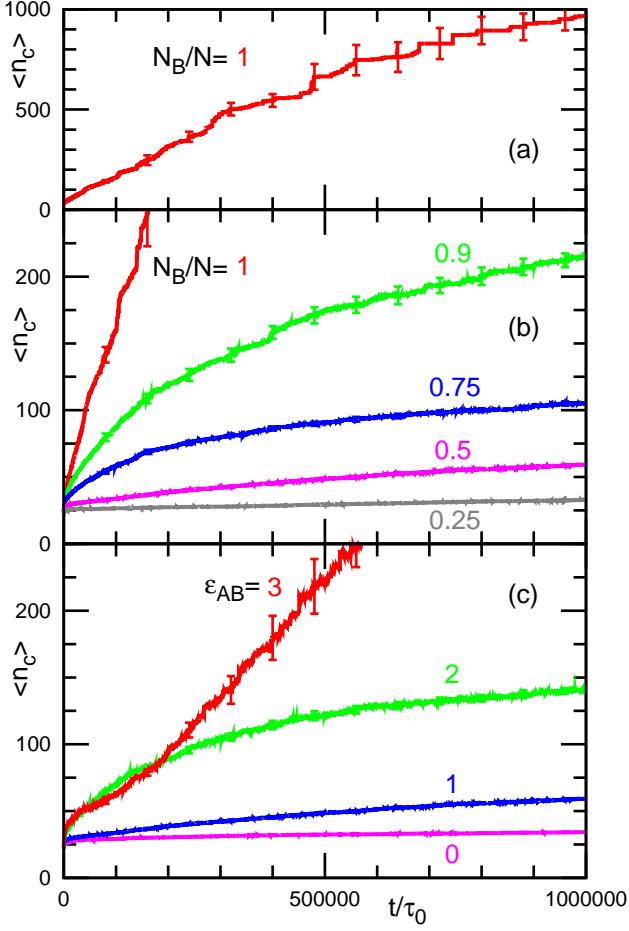}
\caption{\label{fig:ass_cnm}
Time development of the mean cluster size $\langle n_{\rm {cl}} \rangle$.
 (a) $N_{\rm B}/N=1$.
 (b) $N_{\rm B}/N=0.25$, $0.5$, $0.75$, $0.9$, and $1$ at $\varepsilon_{\rm {AB}}=1$.
(c) $\varepsilon_{\rm {AB}}=0$, $1$, $2$, and $3$ at $N_{\rm B}/N=0.5$.
The error-bars are shown at several data points.
}
\end{figure}

\begin{figure}
\includegraphics{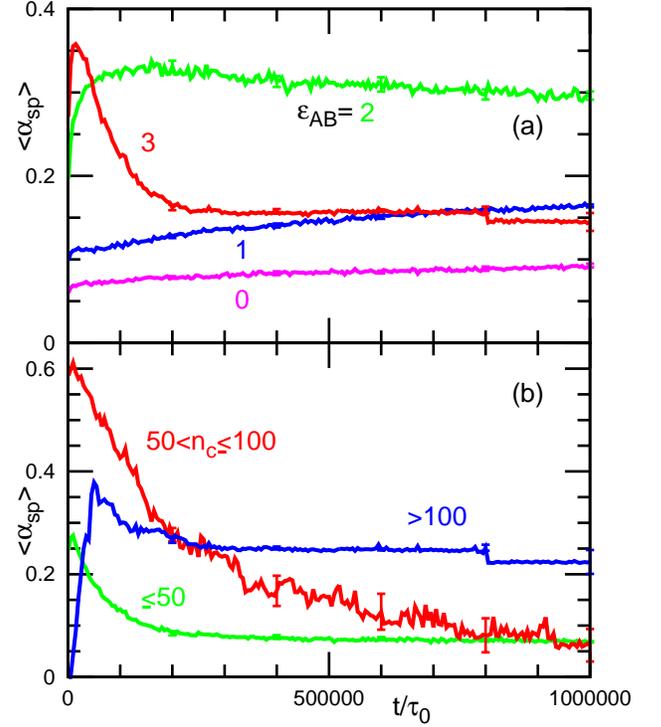}
\caption{\label{fig:ass_asp}
Time development of the mean asphericity $\langle \alpha_{\rm {sp}} \rangle$
of clusters at  $N_{\rm B}/N=0.5$, 
(a) Averaged for all clusters with $n_{\rm {cl}}>10$ at $\varepsilon_{\rm {AB}}=0$, $1$, $2$, and $3$.
(b)  Averaged for clusters with $10<n_{\rm {cl}}\leq 50$, $50<n_{\rm {cl}}\leq 100$, and $n_{\rm {cl}}>100$ 
at $\varepsilon_{\rm {AB}}=3$.
The error-bars are shown at several data points.
}
\end{figure}

Figures \ref{fig:ass_snap_r1}--\ref{fig:ass_asp} show
the self-assembly dynamics of binary mixtures of detergents (type A molecules) and lipids (type B).
First, isolated molecules aggregate into small spherical micelles.
These micelles fuse and grow into disk-like micelles.
In a single-component system ($N=N_{\rm B}$),
large disk-like micelles of size $i_{\rm {cl}} \gtrsim 1000$ close into vesicles via a bowl shape.
In two-component systems,
the type A molecules surround the rim of the bilayer disks of the type B molecules,
and they can stabilize the disk-shaped micelles, bicelles (see Fig. \ref{fig:ass_snap_r1}).
As the number fraction $N_{\rm B}/N$ of the type B molecules or  $\varepsilon_{\rm {AB}}$ increases,
larger bicelles are formed.

Figure~\ref{fig:ass_cnm} shows the time development of 
the mean cluster 
size $\langle n_{\rm {cl}} \rangle$ defined as
\begin{equation}
n_{\rm {cl}}=  \frac{ \sum_{i_{\rm {cl}}=1}^{\infty} i_{\rm {cl}}^2 n_i}
           { \sum_{i_{\rm {cl}}=1}^{\infty} i_{\rm {cl}} n_i},
\end{equation} 
where  $n_i$ is the number of clusters of size $i_{\rm {cl}}$,
and the notation  $\langle ... \rangle$ denotes the average for ten independent simulation runs.
We consider that molecules belong to a
cluster when their distance is less than $r_{\rm {att}}=1.9\sigma$.
For binary mixtures,
the growth of the clusters saturates as the disk rims are covered by the type A molecules
[see Fig. \ref{fig:ass_cnm}(b)].
In the single-component system,
$\langle n_{\rm {cl}} \rangle$  increases linearly with time $t$ 
until vesicle formation starts at $t \simeq 700,000\tau_0$ [see Fig. \ref{fig:ass_cnm}(a)].
This behavior agrees with our previous results obtained by simulations of
the meshless membrane model and by Smoluchowski rate equations \cite{nogu06a}.
In our previous study,
it was found that hydrodynamic interactions increase the rate of aggregation  but do not change the qualitative dynamics of the self-assembly.
Since hydrodynamic interactions  do not change the essence of the dynamics, in this study, we neglect them and use only BD.

As $\varepsilon_{\rm {AB}}$ increases, 
the cluster size increases [see Fig. \ref{fig:ass_cnm}(c)].
This increase is caused by the formation of larger bicelles for  $\varepsilon_{\rm {AB}} \leq 1$
[compare Figs. \ref{fig:ass_snap_r1} and \ref{fig:ass_snap_r02}(a)],
but by separation of the type A domain for $\varepsilon_{\rm {AB}} \geq 2$.
At $\varepsilon_{\rm {AB}} =2$,
the circumferences of the type B disks are not completely surrounded 
by type A molecules [see Fig. \ref{fig:ass_snap_r02}(b)],
so that it is in a partial wetting condition of the type A domains at open edges of the type B domains \cite{dege03,nogu12a}.
At $\varepsilon_{\rm {AB}} =3$,
the small  micelles initially consist of both surfactants; however, the
 two surfactants eventually form completely separated aggregates.
The type A molecules form
spherical micelles and 
the type B molecules form
bilayer disks or vesicles
(see Fig. \ref{fig:ass_snap_r3}).

The shapes of micelles are investigated by calculating
a shape parameter, asphericity $\alpha_{\rm {sp}}$,
which is expressed as \cite{rudn86}
\begin{equation}
\alpha_{\rm {sp}} = \frac{({\lambda_1}-{\lambda_2})^2 + 
  ({\lambda_2}-{\lambda_3})^2+({\lambda_3}-{\lambda_1})^2}{2 (\lambda_1+\lambda_2+\lambda_3)^2},
\end{equation}
where ${\lambda_1} \leq {\lambda_2} \leq {\lambda_3}$ are the 
eigenvalues of the gyration tensor of each micelle. 
The asphericity is the degree of deviation from a spherical 
shape:  $\alpha_{\rm {sp}} = 0$ for spheres, $\alpha_{\rm {sp}}=1$ 
for thin rods, and $\alpha_{\rm {sp}}=0.25$ for thin disks.
The time development of the mean asphericity 
$\langle \alpha_{\rm {sp}} \rangle=  (\sum_{i_{\rm {cl}}} \alpha_{\rm {sp}} i_{\rm {cl}} n_i)
/(\sum_{i_{\rm {cl}}} i_{\rm {cl}} n_i)$ is shown in Fig. \ref{fig:ass_asp}.
For  $\varepsilon_{\rm {AB}} \leq 1$,
the increase in $\langle \alpha_{\rm {sp}} \rangle$ simply
reflects an increase in the cluster size, since
the $\alpha_{\rm {sp}}$ value of each size is almost time-independent:
$\alpha_{\rm {sp}} = 0.1$ (quasi-spherical shapes) and $0.22$ (disks) 
for  $10<n_{\rm {cl}}\leq 50$ and $n_{\rm {cl}}>50$,  respectively.
For  $\varepsilon_{\rm {AB}} \geq 2$,
the cluster shape of each size changes in time [see Fig. \ref{fig:ass_asp}(b)].
Small micelles in the early stage, $t \lesssim 100,000\tau_0$, exhibit
 elongated shapes ($\alpha_{\rm {sp}} \gtrsim 0.5$), 
where one or two spherical type A domains
are connected to a spherical type B domain  [see Figs. \ref{fig:ass_snap_r3}(e) and (f)].
Subsequently, the cluster growth results in the detachment of the type A domains,
as the type A domains become sufficiently large to form stable separated micelles.
This detachment yields a dip in the time evolution curve of the mean cluster size $\langle n_{\rm {cl}} \rangle$
at $t \simeq 100,000\tau_0$.
After the detachment, the type B bilayer disks continue to grow, subsequently resulting in the vesicle formation
[see Figs. \ref{fig:ass_snap_r3}(c) and (d)].

Next, we estimate the bicelle size in a monodisperse distribution.
We assume that a bicelle
 consists of a bilayer disk of the type B domain with radius  $R_{\rm {dis}}-h_0$
and a surrounding ring of the type A domain with radius  $R_{\rm {dis}}$.
The number of type A and  type B molecules in the bicelle are expressed as
$n_{\rm {bc, A}}=4\pi R_{\rm {dis}}/l_0$ and $n_{\rm {bc, B}}=2\pi (R_{\rm {dis}}-h_0)^2/a_0$, respectively,
where $l_0$ and $a_0$ are the circumference length and area per molecule.
The factor $2$ is multiplied to the number of molecules because the membrane has two lipid layers.
In the monodisperse distribution,
the number of molecules are given by
\begin{eqnarray}
n_{\rm {bc, A}} &=& 4\pi \Bigg( \frac{a_0}{l_0^2}\frac{N_{\rm B}}{N_{\rm A}} +\frac{h_0}{l_0} + \sqrt{\Big(\frac{a_0}{l_0^2}\frac{N_{\rm B}}{N_{\rm A}} 
+ \frac{2h_0}{l_0} \Big)\frac{a_0}{l_0^2}\frac{N_{\rm B}}{N_{\rm A}}  } \Bigg)  \nonumber \\
      &\simeq& 8\pi \Big\{ \frac{a_0}{l_0^2}\frac{N_{\rm B}}{N_{\rm A}} +\frac{h_0}{l_0} -\frac{1}{4}\Big(\frac{h_0}{l_0}\Big)^2  \Big\} {\rm \ for\ } \frac{N_{\rm A}}{N_{\rm B}} \ll 1
\end{eqnarray}
and $n_{\rm {bc, B}} = n_{\rm {bc, A}}N_{\rm B}/N_{\rm A}$ for  $a_0/l_0^2 \sim 1$ and $h_0/l_0 \sim 1$.
For $N_{\rm B}/N=0.25$,  $0.5$, $0.75$, and $0.9$ with $a_0= \sigma^2$ and $l_0=h_0= \sigma$,
the total number of molecules in the bicelle are $40$, $90$, $400$, and $2,500$, respectively.
These estimations agree with the simulation results.

\begin{figure}
\includegraphics{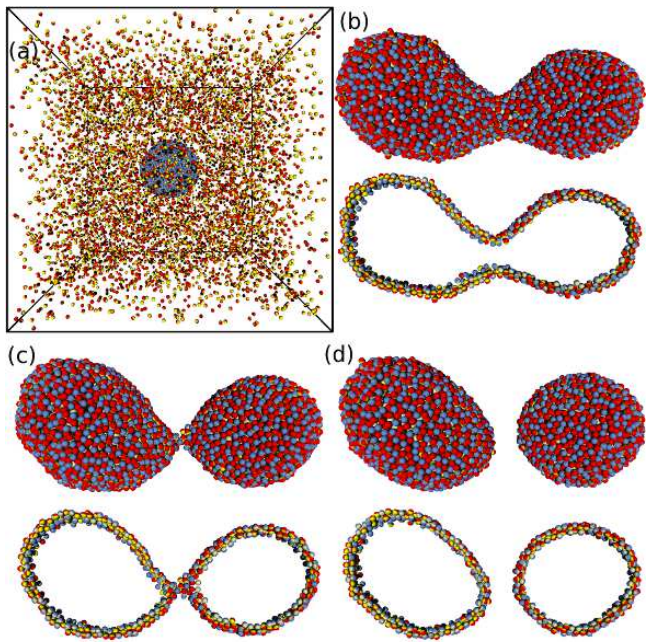}
\caption{\label{fig:fis_snap}
Sequential snapshots of vesicle division due to surfactant adsorption
at $C_{\rm {bd}}^{\rm {A}}=0.8$ and $\varepsilon_{\rm {AB}}=-5$.
(a) $t=0$ (b) $t=8,550\tau_0$. (c) $t=8,563\tau_0$. (d) $t=8,570\tau_0$.
(a) All molecules in the simulation are shown.
(b)-(d) Upper and lower snapshots show whole and cross-sectional-views of vesicles,
respectively.
}
\end{figure}

The bicelle size  varies depending on the initial conditions of the simulations.
When a vesicle is used as the initial state instead of a random gas distribution,
the ruptured vesicle opens up into
a few large bicelles \cite{nogu12a}.
The excess type A molecules form  worm-like micelles or dissolve as isolated molecules.
This scenario is different from the self-assembled bicelles in the present simulations,
where most of the type A molecules belong to bicelles for  $\varepsilon_{\rm {AB}} \leq 1$.
Although the number of detergent molecules in bicelles can be changed via isolated molecules,
the number of lipid molecules can be changed only by fusion and fission of micelles.
Therefore, the growth of the bicelles becomes extremely slow
after stable bicelles, whose rims are completely covered by the detergents, are formed.
The bicelle size is determined more by kinetics 
and depends on the initial states.
Thus, the sample preparation method is important to control the bicelle size.

\begin{figure}
\includegraphics{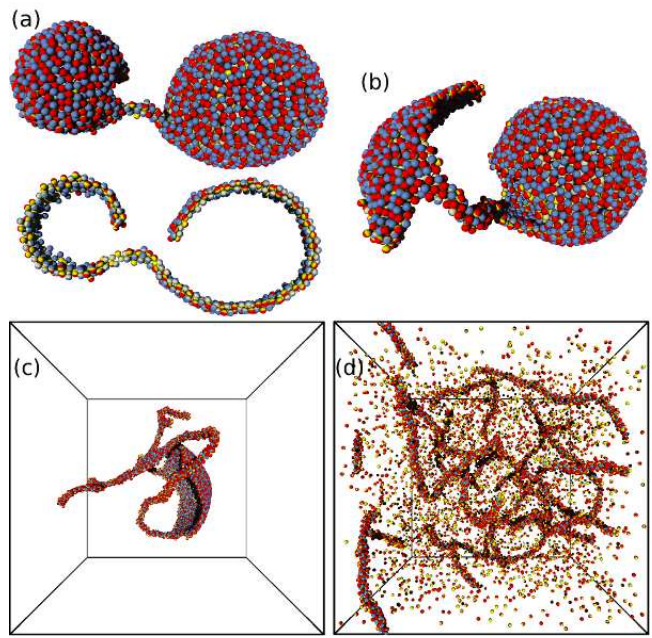}
\caption{\label{fig:worm_snap}
Sequential snapshots of vesicle rupturing into worm-like micelles
at $C_{\rm {bd}}^{\rm {A}}=1.2$ and $\varepsilon_{\rm {AB}}=-5$.
(a) $t=7,600\tau_0$. (b) $t=8,200\tau_0$. (c) $t=9,500\tau_0$. 
(d) $t=30,000\tau_0$.
(a)-(c) Only the largest cluster is shown.
The lower snapshot in (a) shows  the cross-sectional view of the vesicle.
(d) All molecules in the simulation are shown.
}
\end{figure}

\section{Surfactant absorption into vesicles}
\label{sec:abs}

\begin{figure}
\includegraphics{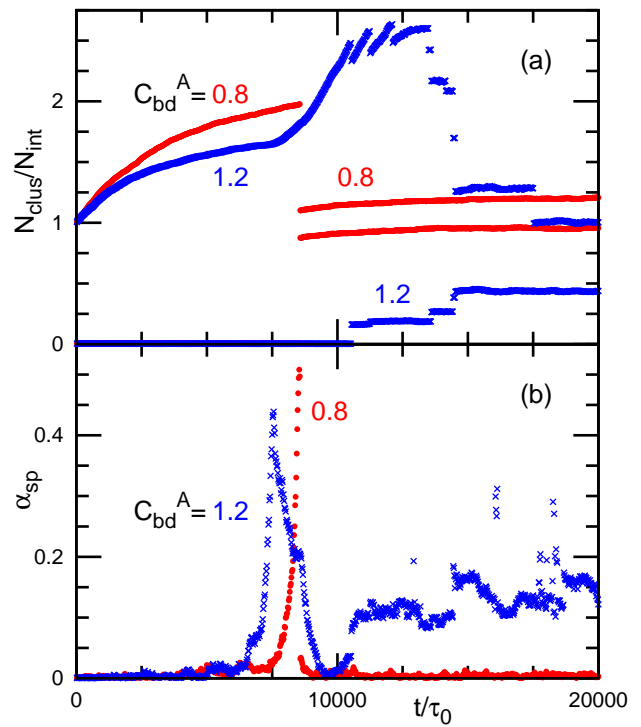}
\caption{\label{fig:cri}
Time development of (a) the  number $N_{\rm {clus}}$ of molecules in the first and second largest clusters
and (b) the aspericity $\alpha_{\rm {sp}}$ of the largest cluster for $C_{\rm {bd}}^{\rm {A}}=0.8$ ($\bullet$) 
and $1.2$ ($\times$).
The cluster size is normalized by the initial vesicle size, $N_{\rm {int}}=2,000$.
The same data are shown in Figs. \ref{fig:fis_snap} and \ref{fig:worm_snap}.
}
\end{figure}

\begin{figure}
\includegraphics{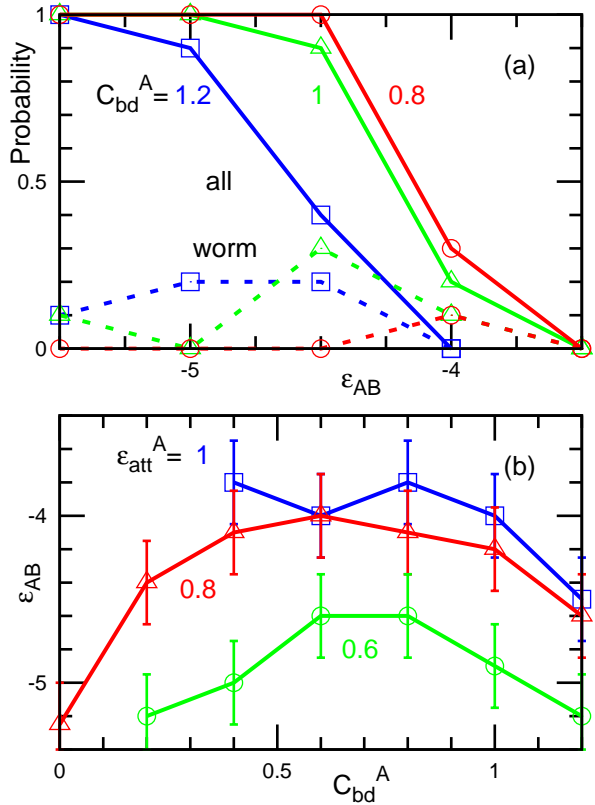}
\caption{\label{fig:fis}
Detergent-adsorption-induced vesicle division and worm-like micelle formation.
(a) Probability of structure changes at $\varepsilon_{\rm att}^{\rm A}=0.8$
for $C_{\rm {bd}}^{\rm {A}}=0.8$ ($\circ$), $1$ ($\triangle$), and $1.2$ ($\Box$).
The solid lines represent the sum of vesicle division and micelle formation.
The dashed lines represent  worm-like micelle formation.
(b) Phase diagram of vesicle structure for
$\varepsilon_{\rm att}^{\rm A}=0.6$ ($\circ$), $0.8$ ($\triangle$), and $1$ ($\Box$).
The vesicle maintains its shape for $\varepsilon_{\rm {AB}}$  values above the solid line.
The vesicle is divided into  two vesicles or transforms into worm-like micelles
below the line.
}
\end{figure}

In this section, we investigate the morphological changes of vesicles that are induced by detergent adsorption.
A lipid vesicle of type B molecules is placed in a solution of type A molecules
[see Fig. \ref{fig:fis_snap}(a)].
The detergent adsorption induces vesicle division into two vesicles or rupture into worm-like micelles
depending on the spontaneous curvature $C_0^{\rm A}$ ($\simeq C_{\rm {bd}}^{\rm A}/2\sigma$) of the type A molecules and the attractive interaction
between different molecules.

\begin{figure}
\includegraphics{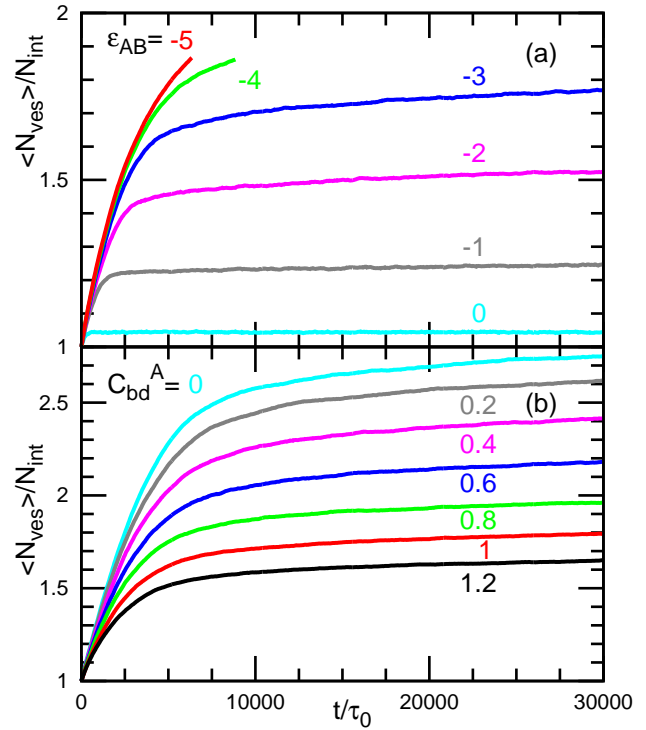}
\caption{\label{fig:cr1}
Time development of the mean number $\langle N_{\rm {ves}} \rangle$ of molecules in the vesicle.
(a) $\varepsilon_{\rm {AB}}=0$, $-1$, $-2$, $-3$, $-4$, and $-5$ at $C_{\rm {bd}}^{\rm {A}}=0.8$.
(b) $C_{\rm {bd}}^{\rm {A}}=0$, $0.2$, $0.4$, $0.6$, $0.8$, $1$, and $1.2$ at $\varepsilon_{\rm {AB}}=-4$.
Divided vesicles are not taken into account for the average in (b).
}
\end{figure}

\begin{figure}
\includegraphics{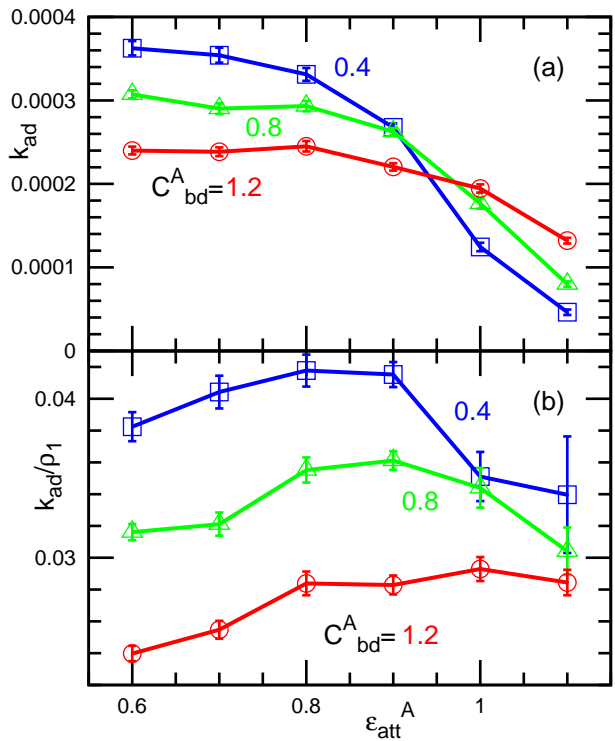}
\caption{\label{fig:slope}
Vesicle initial growth rate $k_{\rm {ad}}=\langle N_{\rm {ves}}\rangle\tau_0/dt|_{t=0}$ 
at $C_{\rm {bd}}^{\rm {A}}=0.4$ ($\Box$), $0.8$ ($\triangle$), $1.2$ ($\circ$) 
and $\varepsilon_{\rm {AB}}=-4$.
The rate is normalized by the initial monomer concentration $\rho_1$ in (b).
}
\end{figure}

Figures \ref{fig:fis_snap}-\ref{fig:cri} show typical examples
of the vesicle division and rupture.
The detergent adsorption first generates an increase in the vesicle size [see Fig. \ref{fig:cri}(a)].
Subsequently the vesicle transforms into a pear-shape, which leads to topological changes [see Fig. \ref{fig:cri}(b)].
The type A molecules are adsorbed onto the outer monolayer (leaflet) of the vesicle,
and subsequently, they slowly move to the inner monolayer.
The resultant asymmetric distribution of the molecules induces
an effective spontaneous curvature $C_0^{\rm {eff}}$ of the bilayer.
The curvature $C_0^{\rm {eff}}$ is caused by two mechanisms.
One is the mismatch between the area difference of the two monolayers
and the number difference of the molecules in two monolayers
as defined in the area difference elasticity (ADE) model \cite{seif97,svet09,saka12}.
The other is the spontaneous curvature of the bilayer, which is induced by 
the asymmetric distribution of type A molecules when $C_0^{\rm A}>0$.
The outer monolayer contains more type A molecules than the inner monolayer.
At $C_0^{\rm A}=0$,
the curvature $C_0^{\rm {eff}}$ is induced only by the area difference,
and the contribution due to the molecular curvature increases with increasing $C_0^{\rm A}$.

For the vesicle division (see Fig. \ref{fig:fis_snap} and corresponding Movie \cite{footnote1}),
parts of the inner monolayer come into contact with each other 
on the inside of the pear neck and subsequently,
the inner monolayer is separated by the outer (trans) monolayer (this saddle-point structure is called the trans-monolayer contact).
The subsequent formation of  a short cylindrical structure called stalk [see Fig. \ref{fig:fis_snap}(c)]
leads to the division of the vesicle.
The two resultant  vesicles can have  sizes equal to that of the original vesicle [see the red lines in Fig. \ref{fig:cri}(a)].
This division pathway is the inverse version of membrane fusion in the modified stalk model \cite{sieg93}.
While this pathway is dominant among the topological changes,
 other dynamics are also at play.
At $C_{\rm {bd}}^{\rm A} \gtrsim 0.8$, the large effective spontaneous curvature $C_0^{\rm {eff}}$ of the bilayer 
causes the opening of a pore at the neck of the pear-shaped vesicle.
This in turn leads to the formation of worm-like micelles or to vesicle division.
The neck transforms into a cylindrical micelle structure and grows to form worm-like micelles.
Other worm-like micelles grow from the edges of the flat bilayer (see Fig. \ref{fig:worm_snap} and corresponding Movie \cite{footnote1}).
After the pore is opened, the surface area for absorption increases (the area of the inner monolayer and the worm-like arms),
so that the absorption rate increases (see Fig. \ref{fig:cri}).
At $C_{\rm {bd}}^{\rm A} \leq 0.6$, no worm-like micelles are formed while
some vesicles transform into worm-like micelles at  $C_{\rm {bd}}^{\rm A} \geq 0.8$ [see Fig. \ref{fig:fis}(a)].
In some simulations at $C_{\rm {bd}}^{\rm A} \geq 0.4$, the opened pores do not expand but lead to pinch-off of the neck 
[the cylindrical connection in the middle of Fig. \ref{fig:worm_snap}(a)], and
subsequently the two pores are closed resulting the formation of two vesicles (occasionally a vesicle and worm-like micelles).
This pore opening is similar to the inverse version of the side-pore (or stalk-bending) fusion pathway \cite{nogu09,nogu01b,nogu02a,muel03},
although the present simulation does not exhibit the stalk intermediate stage during the division.
At  $C_{\rm {bd}}^{\rm A} \leq 0.2$, the neck of the stalk structure often transforms to a bilayer disk 
and is detached as a disk-like micelle.

The number of absorbed detergents increases with increasing attraction between the two types of molecules
and with decreasing $C_{\rm {bd}}^{\rm A}$ (see Fig. \ref{fig:cr1}).
As $\varepsilon_{\rm {AB}}$ decreases from $0$ to $-3$ for $C_{\rm {bd}}^{\rm {A}}=0.8$, 
larger vesicles are formed.
At  $\varepsilon_{\rm {AB}} =-4$,
three of ten vesicles show topological changes 
and the other seven maintain a form of closed vesicles [see Fig. \ref{fig:fis}(a)]. 
The shape transition point is considered as the $\varepsilon_{\rm {AB}}$ value at which
half of the vesicles show either division or rupture.
The transition occurs at weaker levels of attraction for the middle range of $C_{\rm {bd}}^{\rm A} \simeq 0.6$
as shown in Fig. \ref{fig:fis}(b).
This is caused by the competition of two effects.
With increasing  $C_{\rm {bd}}^{\rm A}$,
a lesser number of type A molecules are absorbed into the vesicle but
the bilayer membrane  becomes less stable for the same amount of absorption.
We conclude that the detergents with middle spontaneous curvature $C_0^{\rm A} \simeq 0.3$ ($C_{\rm {bd}}^{\rm A} \simeq 0.6$)
 most efficiently induce  vesicle division.

The initial rate of vesicle growth $k_{\rm {ad}}=\langle N_{\rm {ves}}\rangle\tau_0/dt|_{t=0}$ 
is independent of $\varepsilon_{\rm {AB}}$ when $\varepsilon_{\rm {AB}}<-1$ [see Fig. \ref{fig:cr1}(a)],
since all the detergent molecules making contact with the vesicle are absorbed into the vesicle when the attraction is sufficiently larger than the thermal energy.
The growth rate $k_{\rm {ad}}$ decreases with $\varepsilon_{\rm att}^{\rm A}$ as shown in Fig. \ref{fig:slope}(a).
This is caused by lower CMC values at larger values of $\varepsilon_{\rm att}^{\rm A}$. 
A smaller amount of detergents remains as isolated molecules at larger values of $\varepsilon_{\rm att}^{\rm A}$. 
When the growth rate $k_{\rm {ad}}$ is normalized by the concentration $\rho_1$ of isolated molecules,
it becomes almost constant, as shown in Fig. \ref{fig:slope}(b).
If the concentration is higher than the CMC, the initial adsorption rate would not depend on $\rho_{\rm A}$
while the later evolution dynamics might be modified by $\rho_{\rm A}$.
Here, we consider a fixed concentration of detergents $\rho_{\rm A}$. 
At higher values of $\rho_{\rm A}$, the divided vesicle may exhibit second and third divisions.

Since a solvent-free model is employed, the vesicle can freely change its volume.
 This condition corresponds to high water permeability in comparison to the detergent adsorption rate.
The theoretical study by Svetina and Bo\v{z}i\v{c} \cite{svet09,bozi07} suggests
that higher permeability more easily induces vesicle division.
Our simulation results suggest that the addition of detergent molecules with slightly large head size
efficiently induces vesicle division.

\section{Summary}
\label{sec:sum}

We have studied structure formation dynamics in mixtures of lipids and detergents.
Bicelles are formed by self-assembly from a binary mixture of isolated molecules.
The size of self-assembled bicelles increases with the increasing number fraction of lipids
and repulsion between different molecules.
When a vesicle is used as the an initial simulation state, a large bicelle can be formed,
and the excess amount of the detergent molecules form worm-like micelles.
In contrast, in  self-assembly, most of detergent molecules
are contributed to surround the rims of bicelles.
Thus, the size of the bicelles are  determined by the initial states and assembly kinetics.

When a vesicle is placed in detergent solution,
the detergent adsorption induces vesicle division or vesicle rupture into worm-like micelles.
The vesicle transforms into a pear shape and the contact of the inner monolayers (leaflets) on the inside of the pear neck
induces vesicle division via the stalk intermediate in the modified stalk model.
At large values of spontaneous curvature of the detergent monolayers,
a pore is often opened in the pear neck, and this leads to vesicle division or worm-like micelle formation.
A similar vesicle division has been experimentally observed in multi-component surfactant systems. \cite{tomi11,szos01,hanc04,zepi08,mark10,kuri11,tera12}

While our studies successfully reproduced two types of detergent-induced topological changes,
 other dynamics are observed in experiments such as the inside-out inversion of vesicles. \cite{nomu01}
A pore-opening on the vesicle leads to the formation of a vesicle,  whose outer monolayer is formed by the inner monolayer of the original vesicle.
The  adsorbed detergents form  micelles with lipids in the outer monolayer and
cause removal of the lipids from the outer monolayer,
so that a negative spontaneous curvature of the bilayer is induced.
This process has not been well understood and no simulations on this process have been reported.
Since the size of spherical particles used for both molecules is the same in our simulations,
they do not easily form micelles only from the outer monolayer.
Thus, smaller detergent model molecules are likely necessary to simulate the vesicle inversion.
The mechanisms to determine the assembly and lysis dynamics are not understood completely,
and  open problems remain for further studies.

\begin{acknowledgments}
The computation in this work was partially done using the facilities of the supercomputer Center, 
Institute for Solid State Physics, University of Tokyo.
This study is partially supported by a Grant-in-Aid for Scientific Research 
on Priority Area ``Molecular Science of Fluctuations toward Biological Functions'' from
the Ministry of Education, Culture, Sports, Science, and Technology of Japan.
\end{acknowledgments}


\begin{thebibliography}{64}
\expandafter\ifx\csname natexlab\endcsname\relax\def\natexlab#1{#1}\fi
\expandafter\ifx\csname bibnamefont\endcsname\relax
  \def\bibnamefont#1{#1}\fi
\expandafter\ifx\csname bibfnamefont\endcsname\relax
  \def\bibfnamefont#1{#1}\fi
\expandafter\ifx\csname citenamefont\endcsname\relax
  \def\citenamefont#1{#1}\fi
\expandafter\ifx\csname url\endcsname\relax
  \def\url#1{\texttt{#1}}\fi
\expandafter\ifx\csname urlprefix\endcsname\relax\def\urlprefix{URL }\fi
\providecommand{\bibinfo}[2]{#2}
\providecommand{\eprint}[2][]{\url{#2}}

\bibitem[{\citenamefont{Israelachvili}(2011)}]{isra11}
\bibinfo{author}{\bibfnamefont{J.~N.} \bibnamefont{Israelachvili}},
  \emph{\bibinfo{title}{Intermolecular and surface forces}}
  (\bibinfo{publisher}{Academic Press}, \bibinfo{address}{Burlington, MA},
  \bibinfo{year}{2011}), \bibinfo{edition}{3rd} ed.

\bibitem[{\citenamefont{Safran}(1994)}]{safr94}
\bibinfo{author}{\bibfnamefont{S.~A.} \bibnamefont{Safran}},
  \emph{\bibinfo{title}{Statistical Thermodynamics of Surfaces, Interfaces, and
  Membranes}} (\bibinfo{publisher}{Addison-Wesley}, \bibinfo{address}{Reading,
  MA}, \bibinfo{year}{1994}).

\bibitem[{\citenamefont{Jain and Bates}(2003)}]{jain03}
\bibinfo{author}{\bibfnamefont{S.}~\bibnamefont{Jain}} \bibnamefont{and}
  \bibinfo{author}{\bibfnamefont{F.~S.} \bibnamefont{Bates}},
  \bibinfo{journal}{Science} \textbf{\bibinfo{volume}{300}},
  \bibinfo{pages}{460} (\bibinfo{year}{2003}).

\bibitem[{\citenamefont{Seddon et~al.}(2004)\citenamefont{Seddon, Curnow, and
  Booth}}]{sedd04}
\bibinfo{author}{\bibfnamefont{A.~M.} \bibnamefont{Seddon}},
  \bibinfo{author}{\bibfnamefont{P.}~\bibnamefont{Curnow}}, \bibnamefont{and}
  \bibinfo{author}{\bibfnamefont{P.~J.} \bibnamefont{Booth}},
  \bibinfo{journal}{Biochim.\ Biophys.\ Acta} \textbf{\bibinfo{volume}{1666}},
  \bibinfo{pages}{105} (\bibinfo{year}{2004}).

\bibitem[{\citenamefont{Caffrey}(2009)}]{caff09}
\bibinfo{author}{\bibfnamefont{M.}~\bibnamefont{Caffrey}},
  \bibinfo{journal}{Annu.\ Rev.\ Biophys.} \textbf{\bibinfo{volume}{38}},
  \bibinfo{pages}{29} (\bibinfo{year}{2009}).

\bibitem[{\citenamefont{Ujwal and Bowie}(2011)}]{ujwa11}
\bibinfo{author}{\bibfnamefont{R.}~\bibnamefont{Ujwal}} \bibnamefont{and}
  \bibinfo{author}{\bibfnamefont{J.~U.} \bibnamefont{Bowie}},
  \bibinfo{journal}{Methods} \textbf{\bibinfo{volume}{55}},
  \bibinfo{pages}{337} (\bibinfo{year}{2011}).

\bibitem[{\citenamefont{Walter et~al.}(1991)\citenamefont{Walter, Vinson,
  Kaplun, and Talmon}}]{walt91}
\bibinfo{author}{\bibfnamefont{A.}~\bibnamefont{Walter}},
  \bibinfo{author}{\bibfnamefont{P.~K.} \bibnamefont{Vinson}},
  \bibinfo{author}{\bibfnamefont{A.}~\bibnamefont{Kaplun}}, \bibnamefont{and}
  \bibinfo{author}{\bibfnamefont{Y.}~\bibnamefont{Talmon}},
  \bibinfo{journal}{Biophys.\ J.} \textbf{\bibinfo{volume}{60}},
  \bibinfo{pages}{1315} (\bibinfo{year}{1991}).

\bibitem[{\citenamefont{Yamada et~al.}(2009)\citenamefont{Yamada, Hishida, and
  Torikai}}]{yama09}
\bibinfo{author}{\bibfnamefont{N.~L.} \bibnamefont{Yamada}},
  \bibinfo{author}{\bibfnamefont{M.}~\bibnamefont{Hishida}}, \bibnamefont{and}
  \bibinfo{author}{\bibfnamefont{N.}~\bibnamefont{Torikai}},
  \bibinfo{journal}{Phys.\ Rev.\ E} \textbf{\bibinfo{volume}{79}},
  \bibinfo{pages}{032902} (\bibinfo{year}{2009}).

\bibitem[{\citenamefont{Jain and Bates}(2004)}]{jain04}
\bibinfo{author}{\bibfnamefont{S.}~\bibnamefont{Jain}} \bibnamefont{and}
  \bibinfo{author}{\bibfnamefont{F.~S.} \bibnamefont{Bates}},
  \bibinfo{journal}{Macromol.} \textbf{\bibinfo{volume}{37}},
  \bibinfo{pages}{1511} (\bibinfo{year}{2004}).

\bibitem[{\citenamefont{Christian et~al.}(2009)\citenamefont{Christian, Tian,
  Ellenbroek, Levental, Rajagopal, Janmey, Liu, Baumgart, and
  Discher}}]{chri09}
\bibinfo{author}{\bibfnamefont{D.~A.} \bibnamefont{Christian}},
  \bibinfo{author}{\bibfnamefont{A.}~\bibnamefont{Tian}},
  \bibinfo{author}{\bibfnamefont{W.~G.} \bibnamefont{Ellenbroek}},
  \bibinfo{author}{\bibfnamefont{I.}~\bibnamefont{Levental}},
  \bibinfo{author}{\bibfnamefont{K.}~\bibnamefont{Rajagopal}},
  \bibinfo{author}{\bibfnamefont{P.~A.} \bibnamefont{Janmey}},
  \bibinfo{author}{\bibfnamefont{A.~J.} \bibnamefont{Liu}},
  \bibinfo{author}{\bibfnamefont{T.}~\bibnamefont{Baumgart}}, \bibnamefont{and}
  \bibinfo{author}{\bibfnamefont{D.~E.} \bibnamefont{Discher}},
  \bibinfo{journal}{Nat.\ Mater.} \textbf{\bibinfo{volume}{8}},
  \bibinfo{pages}{843} (\bibinfo{year}{2009}).

\bibitem[{\citenamefont{Sanders and Prosser}(1998)}]{sand98}
\bibinfo{author}{\bibfnamefont{C.~R.} \bibnamefont{Sanders}} \bibnamefont{and}
  \bibinfo{author}{\bibfnamefont{R.~S.} \bibnamefont{Prosser}},
  \bibinfo{journal}{Struct.} \textbf{\bibinfo{volume}{6}},
  \bibinfo{pages}{1227} (\bibinfo{year}{1998}).

\bibitem[{\citenamefont{Angelis and Opella}(2007)}]{ange07}
\bibinfo{author}{\bibfnamefont{A.~A.~D.} \bibnamefont{Angelis}}
  \bibnamefont{and} \bibinfo{author}{\bibfnamefont{S.~J.}
  \bibnamefont{Opella}}, \bibinfo{journal}{Nat.\ Protoc.}
  \textbf{\bibinfo{volume}{2}}, \bibinfo{pages}{2332} (\bibinfo{year}{2007}).

\bibitem[{\citenamefont{Faham and Bowie}(2002)}]{faha02}
\bibinfo{author}{\bibfnamefont{S.}~\bibnamefont{Faham}} \bibnamefont{and}
  \bibinfo{author}{\bibfnamefont{J.~U.} \bibnamefont{Bowie}},
  \bibinfo{journal}{J.\ Mol.\ Biol.} \textbf{\bibinfo{volume}{316}},
  \bibinfo{pages}{1} (\bibinfo{year}{2002}).

\bibitem[{\citenamefont{Leng et~al.}(2002)\citenamefont{Leng, Egelhaaf, and
  Cates}}]{leng02}
\bibinfo{author}{\bibfnamefont{J.}~\bibnamefont{Leng}},
  \bibinfo{author}{\bibfnamefont{S.~U.} \bibnamefont{Egelhaaf}},
  \bibnamefont{and} \bibinfo{author}{\bibfnamefont{M.~E.} \bibnamefont{Cates}},
  \bibinfo{journal}{Europhys.\ Lett.} \textbf{\bibinfo{volume}{59}},
  \bibinfo{pages}{311} (\bibinfo{year}{2002}).

\bibitem[{\citenamefont{Weiss et~al.}(2005)\citenamefont{Weiss, Narayanan,
  Wolf, Gradzielski, Panine, Finet, and Helsby}}]{weis05}
\bibinfo{author}{\bibfnamefont{T.~M.} \bibnamefont{Weiss}},
  \bibinfo{author}{\bibfnamefont{T.}~\bibnamefont{Narayanan}},
  \bibinfo{author}{\bibfnamefont{C.}~\bibnamefont{Wolf}},
  \bibinfo{author}{\bibfnamefont{M.}~\bibnamefont{Gradzielski}},
  \bibinfo{author}{\bibfnamefont{P.}~\bibnamefont{Panine}},
  \bibinfo{author}{\bibfnamefont{S.}~\bibnamefont{Finet}}, \bibnamefont{and}
  \bibinfo{author}{\bibfnamefont{W.~I.} \bibnamefont{Helsby}},
  \bibinfo{journal}{Phys.\ Rev.\ Lett.} \textbf{\bibinfo{volume}{94}},
  \bibinfo{pages}{038303} (\bibinfo{year}{2005}).

\bibitem[{\citenamefont{Bryskhe et~al.}(2005)\citenamefont{Bryskhe, Bulut, and
  Olsson}}]{brys05}
\bibinfo{author}{\bibfnamefont{K.}~\bibnamefont{Bryskhe}},
  \bibinfo{author}{\bibfnamefont{S.}~\bibnamefont{Bulut}}, \bibnamefont{and}
  \bibinfo{author}{\bibfnamefont{U.}~\bibnamefont{Olsson}},
  \bibinfo{journal}{J.\ Phys.\ Chem. B} \textbf{\bibinfo{volume}{109}},
  \bibinfo{pages}{9265} (\bibinfo{year}{2005}).

\bibitem[{\citenamefont{Madenci et~al.}(2011)\citenamefont{Madenci, Salonen,
  Schurtenberger, Pedersen, and Egelhaaf}}]{made11}
\bibinfo{author}{\bibfnamefont{D.}~\bibnamefont{Madenci}},
  \bibinfo{author}{\bibfnamefont{A.}~\bibnamefont{Salonen}},
  \bibinfo{author}{\bibfnamefont{P.}~\bibnamefont{Schurtenberger}},
  \bibinfo{author}{\bibfnamefont{J.~S.} \bibnamefont{Pedersen}},
  \bibnamefont{and} \bibinfo{author}{\bibfnamefont{S.~U.}
  \bibnamefont{Egelhaaf}}, \bibinfo{journal}{Phys.\ Chem.\ Chem.\ Phys.}
  \textbf{\bibinfo{volume}{13}}, \bibinfo{pages}{3171} (\bibinfo{year}{2011}).

\bibitem[{\citenamefont{Gummel et~al.}(2011)\citenamefont{Gummel, Sztucki,
  Narayanan, and Gradzielski}}]{gumm11}
\bibinfo{author}{\bibfnamefont{J.}~\bibnamefont{Gummel}},
  \bibinfo{author}{\bibfnamefont{M.}~\bibnamefont{Sztucki}},
  \bibinfo{author}{\bibfnamefont{T.}~\bibnamefont{Narayanan}},
  \bibnamefont{and}
  \bibinfo{author}{\bibfnamefont{M.}~\bibnamefont{Gradzielski}},
  \bibinfo{journal}{Soft\ Matter} \textbf{\bibinfo{volume}{7}},
  \bibinfo{pages}{5731} (\bibinfo{year}{2011}).

\bibitem[{\citenamefont{Noguchi and Gompper}(2006{\natexlab{a}})}]{nogu06a}
\bibinfo{author}{\bibfnamefont{H.}~\bibnamefont{Noguchi}} \bibnamefont{and}
  \bibinfo{author}{\bibfnamefont{G.}~\bibnamefont{Gompper}},
  \bibinfo{journal}{J.\ Chem.\ Phys.} \textbf{\bibinfo{volume}{125}},
  \bibinfo{pages}{164908} (\bibinfo{year}{2006}{\natexlab{a}}).

\bibitem[{\citenamefont{Nomura et~al.}(2001)\citenamefont{Nomura, Nagata,
  Inaba, Hiramatsu, Hotani, and Takiguchi}}]{nomu01}
\bibinfo{author}{\bibfnamefont{F.}~\bibnamefont{Nomura}},
  \bibinfo{author}{\bibfnamefont{M.}~\bibnamefont{Nagata}},
  \bibinfo{author}{\bibfnamefont{T.}~\bibnamefont{Inaba}},
  \bibinfo{author}{\bibfnamefont{H.}~\bibnamefont{Hiramatsu}},
  \bibinfo{author}{\bibfnamefont{H.}~\bibnamefont{Hotani}}, \bibnamefont{and}
  \bibinfo{author}{\bibfnamefont{K.}~\bibnamefont{Takiguchi}},
  \bibinfo{journal}{Proc.\ Natl.\ Acad.\ Sci.\ USA}
  \textbf{\bibinfo{volume}{98}}, \bibinfo{pages}{2340} (\bibinfo{year}{2001}).

\bibitem[{\citenamefont{Staneva et~al.}(2005)\citenamefont{Staneva, Seigneuret,
  Koumanov, Trugnan, and Angelova}}]{stan05}
\bibinfo{author}{\bibfnamefont{G.}~\bibnamefont{Staneva}},
  \bibinfo{author}{\bibfnamefont{M.}~\bibnamefont{Seigneuret}},
  \bibinfo{author}{\bibfnamefont{K.}~\bibnamefont{Koumanov}},
  \bibinfo{author}{\bibfnamefont{G.}~\bibnamefont{Trugnan}}, \bibnamefont{and}
  \bibinfo{author}{\bibfnamefont{M.~I.} \bibnamefont{Angelova}},
  \bibinfo{journal}{Chem.\ Phys.\ Lipids} \textbf{\bibinfo{volume}{136}},
  \bibinfo{pages}{55} (\bibinfo{year}{2005}).

\bibitem[{\citenamefont{Sudbrack et~al.}(2011)\citenamefont{Sudbrack, Archilha,
  Itri, and Riske}}]{sudb11}
\bibinfo{author}{\bibfnamefont{T.~P.} \bibnamefont{Sudbrack}},
  \bibinfo{author}{\bibfnamefont{N.~L.} \bibnamefont{Archilha}},
  \bibinfo{author}{\bibfnamefont{R.}~\bibnamefont{Itri}}, \bibnamefont{and}
  \bibinfo{author}{\bibfnamefont{K.~A.} \bibnamefont{Riske}},
  \bibinfo{journal}{J.\ Phys.\ Chem.\ B} \textbf{\bibinfo{volume}{115}},
  \bibinfo{pages}{269} (\bibinfo{year}{2011}).

\bibitem[{\citenamefont{Tomita et~al.}(2011)\citenamefont{Tomita, Sugawara, and
  Wakamoto}}]{tomi11}
\bibinfo{author}{\bibfnamefont{T.}~\bibnamefont{Tomita}},
  \bibinfo{author}{\bibfnamefont{T.}~\bibnamefont{Sugawara}}, \bibnamefont{and}
  \bibinfo{author}{\bibfnamefont{Y.}~\bibnamefont{Wakamoto}},
  \bibinfo{journal}{Langmuir} \textbf{\bibinfo{volume}{27}},
  \bibinfo{pages}{10106} (\bibinfo{year}{2011}).

\bibitem[{\citenamefont{Elsayed and Cevc}(2011)}]{elsa11}
\bibinfo{author}{\bibfnamefont{M.~M.~A.} \bibnamefont{Elsayed}}
  \bibnamefont{and} \bibinfo{author}{\bibfnamefont{G.}~\bibnamefont{Cevc}},
  \bibinfo{journal}{Biochim.\ Biophys.\ Acta} \textbf{\bibinfo{volume}{1808}},
  \bibinfo{pages}{140} (\bibinfo{year}{2011}).

\bibitem[{\citenamefont{Hamada et~al.}(2012)\citenamefont{Hamada, Hagihara,
  Morita, Vestergaard, Tsujino, and Takagi}}]{hama12}
\bibinfo{author}{\bibfnamefont{T.}~\bibnamefont{Hamada}},
  \bibinfo{author}{\bibfnamefont{H.}~\bibnamefont{Hagihara}},
  \bibinfo{author}{\bibfnamefont{M.}~\bibnamefont{Morita}},
  \bibinfo{author}{\bibfnamefont{M.~C.} \bibnamefont{Vestergaard}},
  \bibinfo{author}{\bibfnamefont{Y.}~\bibnamefont{Tsujino}}, \bibnamefont{and}
  \bibinfo{author}{\bibfnamefont{M.}~\bibnamefont{Takagi}},
  \bibinfo{journal}{J.\ Phys.\ Chem.\ Lett.} \textbf{\bibinfo{volume}{3}},
  \bibinfo{pages}{430} (\bibinfo{year}{2012}).

\bibitem[{\citenamefont{Szostak et~al.}(2001)\citenamefont{Szostak, Bartel, and
  Luisi}}]{szos01}
\bibinfo{author}{\bibfnamefont{J.~W.} \bibnamefont{Szostak}},
  \bibinfo{author}{\bibfnamefont{D.~P.} \bibnamefont{Bartel}},
  \bibnamefont{and} \bibinfo{author}{\bibfnamefont{P.~L.} \bibnamefont{Luisi}},
  \bibinfo{journal}{Nature} \textbf{\bibinfo{volume}{409}},
  \bibinfo{pages}{387} (\bibinfo{year}{2001}).

\bibitem[{\citenamefont{Hanczyc and Szostak}(2004)}]{hanc04}
\bibinfo{author}{\bibfnamefont{M.~M.} \bibnamefont{Hanczyc}} \bibnamefont{and}
  \bibinfo{author}{\bibfnamefont{J.~W.} \bibnamefont{Szostak}},
  \bibinfo{journal}{Curr.\ Opin.\ Chem.\ Biol.} \textbf{\bibinfo{volume}{8}},
  \bibinfo{pages}{660} (\bibinfo{year}{2004}).

\bibitem[{\citenamefont{Zepik and Walde}(2008)}]{zepi08}
\bibinfo{author}{\bibfnamefont{H.~H.} \bibnamefont{Zepik}} \bibnamefont{and}
  \bibinfo{author}{\bibfnamefont{P.}~\bibnamefont{Walde}},
  \bibinfo{journal}{ChemBioChem} \textbf{\bibinfo{volume}{9}},
  \bibinfo{pages}{2771} (\bibinfo{year}{2008}).

\bibitem[{\citenamefont{Kurihara et~al.}(2011)\citenamefont{Kurihara, Tamura,
  Shohda, Toyota, Suzuki, and Sugawara}}]{kuri11}
\bibinfo{author}{\bibfnamefont{K.}~\bibnamefont{Kurihara}},
  \bibinfo{author}{\bibfnamefont{M.}~\bibnamefont{Tamura}},
  \bibinfo{author}{\bibfnamefont{K.}~\bibnamefont{Shohda}},
  \bibinfo{author}{\bibfnamefont{T.}~\bibnamefont{Toyota}},
  \bibinfo{author}{\bibfnamefont{K.}~\bibnamefont{Suzuki}}, \bibnamefont{and}
  \bibinfo{author}{\bibfnamefont{T.}~\bibnamefont{Sugawara}},
  \bibinfo{journal}{Nat.\ Chem.} \textbf{\bibinfo{volume}{3}},
  \bibinfo{pages}{775} (\bibinfo{year}{2011}).

\bibitem[{\citenamefont{Terasawa et~al.}(2012)\citenamefont{Terasawa,
  Nishimura, Suzuki, Matsuura, and Yomo}}]{tera12}
\bibinfo{author}{\bibfnamefont{H.}~\bibnamefont{Terasawa}},
  \bibinfo{author}{\bibfnamefont{K.}~\bibnamefont{Nishimura}},
  \bibinfo{author}{\bibfnamefont{H.}~\bibnamefont{Suzuki}},
  \bibinfo{author}{\bibfnamefont{T.}~\bibnamefont{Matsuura}}, \bibnamefont{and}
  \bibinfo{author}{\bibfnamefont{T.}~\bibnamefont{Yomo}},
  \bibinfo{journal}{Proc.\ Natl.\ Acad.\ Sci.\ USA}
  \textbf{\bibinfo{volume}{109}}, \bibinfo{pages}{5942} (\bibinfo{year}{2012}).

\bibitem[{\citenamefont{Svetina}(2009)}]{svet09}
\bibinfo{author}{\bibfnamefont{S.}~\bibnamefont{Svetina}},
  \bibinfo{journal}{ChemPhysChem} \textbf{\bibinfo{volume}{10}},
  \bibinfo{pages}{2769} (\bibinfo{year}{2009}).

\bibitem[{\citenamefont{Bo\v{z}i\v{c} and Svetina}(2007)}]{bozi07}
\bibinfo{author}{\bibfnamefont{B.}~\bibnamefont{Bo\v{z}i\v{c}}}
  \bibnamefont{and} \bibinfo{author}{\bibfnamefont{S.}~\bibnamefont{Svetina}},
  \bibinfo{journal}{Eur.\ Phys.\ J.\ E} \textbf{\bibinfo{volume}{24}},
  \bibinfo{pages}{79} (\bibinfo{year}{2007}).

\bibitem[{\citenamefont{Markvoort et~al.}(2010)\citenamefont{Markvoort,
  Pfleger, Staffhorst, Hilbers, {van Santen}, Killian, and {de
  Kruijff}}}]{mark10}
\bibinfo{author}{\bibfnamefont{A.~J.} \bibnamefont{Markvoort}},
  \bibinfo{author}{\bibfnamefont{N.}~\bibnamefont{Pfleger}},
  \bibinfo{author}{\bibfnamefont{R.}~\bibnamefont{Staffhorst}},
  \bibinfo{author}{\bibfnamefont{P.~A.~J.} \bibnamefont{Hilbers}},
  \bibinfo{author}{\bibfnamefont{R.~A.} \bibnamefont{{van Santen}}},
  \bibinfo{author}{\bibfnamefont{J.~A.} \bibnamefont{Killian}},
  \bibnamefont{and} \bibinfo{author}{\bibfnamefont{B.}~\bibnamefont{{de
  Kruijff}}}, \bibinfo{journal}{Biophys.\ J.} \textbf{\bibinfo{volume}{99}},
  \bibinfo{pages}{1520} (\bibinfo{year}{2010}).

\bibitem[{\citenamefont{Jahn and Grubm{\"u}ller}(2002)}]{jahn02}
\bibinfo{author}{\bibfnamefont{R.}~\bibnamefont{Jahn}} \bibnamefont{and}
  \bibinfo{author}{\bibfnamefont{H.}~\bibnamefont{Grubm{\"u}ller}},
  \bibinfo{journal}{Curr.\ Opin.\ Cell\ Biol.} \textbf{\bibinfo{volume}{14}},
  \bibinfo{pages}{488} (\bibinfo{year}{2002}).

\bibitem[{\citenamefont{Chernomordik and Kozlov}(2008)}]{cher08}
\bibinfo{author}{\bibfnamefont{L.~V.} \bibnamefont{Chernomordik}}
  \bibnamefont{and} \bibinfo{author}{\bibfnamefont{M.~M.}
  \bibnamefont{Kozlov}}, \bibinfo{journal}{Nat.\ Struct.\ Mol.\ Biol.}
  \textbf{\bibinfo{volume}{15}}, \bibinfo{pages}{675} (\bibinfo{year}{2008}).

\bibitem[{\citenamefont{Nikolaus et~al.}(2011)\citenamefont{Nikolaus, Warner,
  O'Shaughnessy, and Herrmann}}]{niko11}
\bibinfo{author}{\bibfnamefont{J.}~\bibnamefont{Nikolaus}},
  \bibinfo{author}{\bibfnamefont{J.~M.} \bibnamefont{Warner}},
  \bibinfo{author}{\bibfnamefont{B.}~\bibnamefont{O'Shaughnessy}},
  \bibnamefont{and} \bibinfo{author}{\bibfnamefont{A.}~\bibnamefont{Herrmann}},
  \bibinfo{journal}{Curr.\ Top.\ Membr.} \textbf{\bibinfo{volume}{68}},
  \bibinfo{pages}{1} (\bibinfo{year}{2011}).

\bibitem[{\citenamefont{Markvoort and Marrink}(2011)}]{mark11}
\bibinfo{author}{\bibfnamefont{A.~J.} \bibnamefont{Markvoort}}
  \bibnamefont{and} \bibinfo{author}{\bibfnamefont{S.~J.}
  \bibnamefont{Marrink}}, \bibinfo{journal}{Curr.\ Top.\ Membr.}
  \textbf{\bibinfo{volume}{68}}, \bibinfo{pages}{259} (\bibinfo{year}{2011}).

\bibitem[{\citenamefont{M{\"u}ller and Schick}(2011)}]{mull11}
\bibinfo{author}{\bibfnamefont{M.}~\bibnamefont{M{\"u}ller}} \bibnamefont{and}
  \bibinfo{author}{\bibfnamefont{M.}~\bibnamefont{Schick}},
  \bibinfo{journal}{Curr.\ Top.\ Membr.} \textbf{\bibinfo{volume}{68}},
  \bibinfo{pages}{295} (\bibinfo{year}{2011}).

\bibitem[{\citenamefont{Noguchi and Takasu}(2002{\natexlab{a}})}]{nogu02a}
\bibinfo{author}{\bibfnamefont{H.}~\bibnamefont{Noguchi}} \bibnamefont{and}
  \bibinfo{author}{\bibfnamefont{M.}~\bibnamefont{Takasu}},
  \bibinfo{journal}{Biophys.\ J.} \textbf{\bibinfo{volume}{83}},
  \bibinfo{pages}{299} (\bibinfo{year}{2002}{\natexlab{a}}).

\bibitem[{\citenamefont{Noguchi and Takasu}(2002{\natexlab{b}})}]{nogu02b}
\bibinfo{author}{\bibfnamefont{H.}~\bibnamefont{Noguchi}} \bibnamefont{and}
  \bibinfo{author}{\bibfnamefont{M.}~\bibnamefont{Takasu}},
  \bibinfo{journal}{Phys.\ Rev.\ E} \textbf{\bibinfo{volume}{65}},
  \bibinfo{pages}{051907} (\bibinfo{year}{2002}{\natexlab{b}}).

\bibitem[{\citenamefont{Noguchi}(2003)}]{nogu03}
\bibinfo{author}{\bibfnamefont{H.}~\bibnamefont{Noguchi}},
  \bibinfo{journal}{Phys.\ Rev.\ E} \textbf{\bibinfo{volume}{67}},
  \bibinfo{pages}{041901} (\bibinfo{year}{2003}).

\bibitem[{\citenamefont{Yamamoto and Hyodo}(2003)}]{yama03}
\bibinfo{author}{\bibfnamefont{S.}~\bibnamefont{Yamamoto}} \bibnamefont{and}
  \bibinfo{author}{\bibfnamefont{S.~A.} \bibnamefont{Hyodo}},
  \bibinfo{journal}{J.\ Chem.\ Phys.} \textbf{\bibinfo{volume}{118}},
  \bibinfo{pages}{7937} (\bibinfo{year}{2003}).

\bibitem[{\citenamefont{Smith and Uspal}(2007)}]{smit07}
\bibinfo{author}{\bibfnamefont{K.~A.} \bibnamefont{Smith}} \bibnamefont{and}
  \bibinfo{author}{\bibfnamefont{W.~E.} \bibnamefont{Uspal}},
  \bibinfo{journal}{J.\ Chem.\ Phys.} \textbf{\bibinfo{volume}{126}},
  \bibinfo{pages}{075102} (\bibinfo{year}{2007}).

\bibitem[{\citenamefont{Siegel}(1993)}]{sieg93}
\bibinfo{author}{\bibfnamefont{D.~P.} \bibnamefont{Siegel}},
  \bibinfo{journal}{Biophys.\ J.} \textbf{\bibinfo{volume}{65}},
  \bibinfo{pages}{2124} (\bibinfo{year}{1993}).

\bibitem[{\citenamefont{M{\"u}ller et~al.}(2006)\citenamefont{M{\"u}ller,
  Katsov, and Schick}}]{muel06}
\bibinfo{author}{\bibfnamefont{M.}~\bibnamefont{M{\"u}ller}},
  \bibinfo{author}{\bibfnamefont{K.}~\bibnamefont{Katsov}}, \bibnamefont{and}
  \bibinfo{author}{\bibfnamefont{M.}~\bibnamefont{Schick}},
  \bibinfo{journal}{Phys.\ Rep.} \textbf{\bibinfo{volume}{434}},
  \bibinfo{pages}{113} (\bibinfo{year}{2006}).

\bibitem[{\citenamefont{Venturoli et~al.}(2006)\citenamefont{Venturoli,
  Sperotto, Kranenburg, and Smit}}]{vent06}
\bibinfo{author}{\bibfnamefont{M.}~\bibnamefont{Venturoli}},
  \bibinfo{author}{\bibfnamefont{M.~M.} \bibnamefont{Sperotto}},
  \bibinfo{author}{\bibfnamefont{M.}~\bibnamefont{Kranenburg}},
  \bibnamefont{and} \bibinfo{author}{\bibfnamefont{B.}~\bibnamefont{Smit}},
  \bibinfo{journal}{Phys.\ Rep.} \textbf{\bibinfo{volume}{437}},
  \bibinfo{pages}{1} (\bibinfo{year}{2006}).

\bibitem[{\citenamefont{Noguchi}(2009)}]{nogu09}
\bibinfo{author}{\bibfnamefont{H.}~\bibnamefont{Noguchi}},
  \bibinfo{journal}{J.\ Phys.\ Soc.\ Jpn.} \textbf{\bibinfo{volume}{78}},
  \bibinfo{pages}{041007} (\bibinfo{year}{2009}).

\bibitem[{\citenamefont{Marrink et~al.}(2009)\citenamefont{Marrink, de\ Vries,
  and Tieleman}}]{marr09}
\bibinfo{author}{\bibfnamefont{S.~J.} \bibnamefont{Marrink}},
  \bibinfo{author}{\bibfnamefont{A.~H.} \bibnamefont{de\ Vries}},
  \bibnamefont{and} \bibinfo{author}{\bibfnamefont{D.~P.}
  \bibnamefont{Tieleman}}, \bibinfo{journal}{Biochim.\ Biophys.\ Acta}
  \textbf{\bibinfo{volume}{1788}}, \bibinfo{pages}{149} (\bibinfo{year}{2009}).

\bibitem[{\citenamefont{Shinoda et~al.}(2012)\citenamefont{Shinoda, DeVane, and
  Klein}}]{shin12}
\bibinfo{author}{\bibfnamefont{W.}~\bibnamefont{Shinoda}},
  \bibinfo{author}{\bibfnamefont{R.}~\bibnamefont{DeVane}}, \bibnamefont{and}
  \bibinfo{author}{\bibfnamefont{M.~L.} \bibnamefont{Klein}},
  \bibinfo{journal}{Curr.\ Opin.\ Struct.\ Biol.}
  \textbf{\bibinfo{volume}{22}}, \bibinfo{pages}{175} (\bibinfo{year}{2012}).

\bibitem[{\citenamefont{Noguchi}(2011)}]{nogu11}
\bibinfo{author}{\bibfnamefont{H.}~\bibnamefont{Noguchi}},
  \bibinfo{journal}{J.\ Chem.\ Phys.} \textbf{\bibinfo{volume}{134}},
  \bibinfo{pages}{055101} (\bibinfo{year}{2011}).

\bibitem[{\citenamefont{Noguchi}(2012{\natexlab{a}})}]{nogu12a}
\bibinfo{author}{\bibfnamefont{H.}~\bibnamefont{Noguchi}},
  \bibinfo{journal}{Soft Matter} \textbf{\bibinfo{volume}{8}},
  \bibinfo{pages}{8926} (\bibinfo{year}{2012}{\natexlab{a}}).

\bibitem[{\citenamefont{Noguchi and Gompper}(2006{\natexlab{b}})}]{nogu06}
\bibinfo{author}{\bibfnamefont{H.}~\bibnamefont{Noguchi}} \bibnamefont{and}
  \bibinfo{author}{\bibfnamefont{G.}~\bibnamefont{Gompper}},
  \bibinfo{journal}{Phys.\ Rev.\ E} \textbf{\bibinfo{volume}{73}},
  \bibinfo{pages}{021903} (\bibinfo{year}{2006}{\natexlab{b}}).

\bibitem[{\citenamefont{Hamm and Kozlov}(1998)}]{hamm98}
\bibinfo{author}{\bibfnamefont{M.}~\bibnamefont{Hamm}} \bibnamefont{and}
  \bibinfo{author}{\bibfnamefont{M.~M.} \bibnamefont{Kozlov}},
  \bibinfo{journal}{Eur.\ Phys.\ J.\ B} \textbf{\bibinfo{volume}{6}},
  \bibinfo{pages}{519} (\bibinfo{year}{1998}).

\bibitem[{\citenamefont{Hamm and Kozlov}(2000)}]{hamm00}
\bibinfo{author}{\bibfnamefont{M.}~\bibnamefont{Hamm}} \bibnamefont{and}
  \bibinfo{author}{\bibfnamefont{M.~M.} \bibnamefont{Kozlov}},
  \bibinfo{journal}{Eur.\ Phys.\ J.\ E} \textbf{\bibinfo{volume}{3}},
  \bibinfo{pages}{323} (\bibinfo{year}{2000}).

\bibitem[{\citenamefont{Shiba and Noguchi}(2011)}]{shiba11}
\bibinfo{author}{\bibfnamefont{H.}~\bibnamefont{Shiba}} \bibnamefont{and}
  \bibinfo{author}{\bibfnamefont{H.}~\bibnamefont{Noguchi}},
  \bibinfo{journal}{Phys. Rev. E} \textbf{\bibinfo{volume}{84}},
  \bibinfo{pages}{031926} (\bibinfo{year}{2011}).

\bibitem[{\citenamefont{Noguchi}(2012{\natexlab{b}})}]{nogu12}
\bibinfo{author}{\bibfnamefont{H.}~\bibnamefont{Noguchi}},
  \bibinfo{journal}{Soft Matter} \textbf{\bibinfo{volume}{8}},
  \bibinfo{pages}{3146} (\bibinfo{year}{2012}{\natexlab{b}}).

\bibitem[{\citenamefont{Wu et~al.}(1977)\citenamefont{Wu, Jacobson, and
  Papahadjopoulos}}]{wu77}
\bibinfo{author}{\bibfnamefont{E.~S.} \bibnamefont{Wu}},
  \bibinfo{author}{\bibfnamefont{K.}~\bibnamefont{Jacobson}}, \bibnamefont{and}
  \bibinfo{author}{\bibfnamefont{D.}~\bibnamefont{Papahadjopoulos}},
  \bibinfo{journal}{Biochemistry} \textbf{\bibinfo{volume}{16}},
  \bibinfo{pages}{3936} (\bibinfo{year}{1977}).

\bibitem[{\citenamefont{de~Gennes et~al.}(2003)\citenamefont{de~Gennes,
  Brochard-Wyart, and Quere}}]{dege03}
\bibinfo{author}{\bibfnamefont{P.~G.} \bibnamefont{de~Gennes}},
  \bibinfo{author}{\bibfnamefont{F.}~\bibnamefont{Brochard-Wyart}},
  \bibnamefont{and} \bibinfo{author}{\bibfnamefont{D.}~\bibnamefont{Quere}},
  \emph{\bibinfo{title}{Capillarity and wetting phenomena: Drops, bubbles,
  perls, waves}} (\bibinfo{publisher}{Springer}, \bibinfo{address}{New York},
  \bibinfo{year}{2003}).

\bibitem[{\citenamefont{Rudnick and Gaspari}(1986)}]{rudn86}
\bibinfo{author}{\bibfnamefont{J.}~\bibnamefont{Rudnick}} \bibnamefont{and}
  \bibinfo{author}{\bibfnamefont{G.}~\bibnamefont{Gaspari}},
  \bibinfo{journal}{J. Phys.\ A:\ Math.\ Gen.} \textbf{\bibinfo{volume}{19}},
  \bibinfo{pages}{L191} (\bibinfo{year}{1986}).

\bibitem[{\citenamefont{Seifert}(1997)}]{seif97}
\bibinfo{author}{\bibfnamefont{U.}~\bibnamefont{Seifert}},
  \bibinfo{journal}{Adv.\ Phys.} \textbf{\bibinfo{volume}{46}},
  \bibinfo{pages}{13} (\bibinfo{year}{1997}).

\bibitem[{\citenamefont{Sakashita et~al.}(2012)\citenamefont{Sakashita,
  Urakami, Ziherl, and Imai}}]{saka12}
\bibinfo{author}{\bibfnamefont{A.}~\bibnamefont{Sakashita}},
  \bibinfo{author}{\bibfnamefont{N.}~\bibnamefont{Urakami}},
  \bibinfo{author}{\bibfnamefont{P.}~\bibnamefont{Ziherl}}, \bibnamefont{and}
  \bibinfo{author}{\bibfnamefont{M.}~\bibnamefont{Imai}},
  \bibinfo{journal}{Soft\ Matter} \textbf{\bibinfo{volume}{8}},
  \bibinfo{pages}{8569} (\bibinfo{year}{2012}).

\bibitem[{foo()}]{footnote1}
\bibinfo{note}{See supplementary material at [XXX] for movies of vesicle
  division and worm-like micelle formation.}

\bibitem[{\citenamefont{Noguchi and Takasu}(2001)}]{nogu01b}
\bibinfo{author}{\bibfnamefont{H.}~\bibnamefont{Noguchi}} \bibnamefont{and}
  \bibinfo{author}{\bibfnamefont{M.}~\bibnamefont{Takasu}},
  \bibinfo{journal}{J.\ Chem.\ Phys.} \textbf{\bibinfo{volume}{115}},
  \bibinfo{pages}{9547} (\bibinfo{year}{2001}).

\bibitem[{\citenamefont{M{\"u}ller et~al.}(2003)\citenamefont{M{\"u}ller,
  Katsov, and Schick}}]{muel03}
\bibinfo{author}{\bibfnamefont{M.}~\bibnamefont{M{\"u}ller}},
  \bibinfo{author}{\bibfnamefont{K.}~\bibnamefont{Katsov}}, \bibnamefont{and}
  \bibinfo{author}{\bibfnamefont{M.}~\bibnamefont{Schick}},
  \bibinfo{journal}{Biophys.\ J.} \textbf{\bibinfo{volume}{85}},
  \bibinfo{pages}{1611} (\bibinfo{year}{2003}).

\end{thebibliography}

\end{document}